\def\preprint{0}                %  submitted version
\def\preprint{1}                %  preprint
\def\comment#1{}
\if\preprint1
\documentclass[12pt,preprint]{aastex}
        \usepackage{times}
        \usepackage{calc}

\else
        \documentstyle[astrop-bib,referee,times]{mn}
        \newcommand{\includegraphics}[1]{}
\fi

\def\oversim#1#2{\lower0.5pt\vbox{\baselineskip0pt \lineskip-0.5pt
     \ialign{$\mathsurround0pt #1\hfil##\hfil$\crcr#2\crcr\sim\crcr}}}
\shorttitle{The evolution of NGC 7027}
\shortauthors{Zijlstra, van Hoof, \&\ Perley}

\begin{document}

%\title{The evolving radio spectrum of NGC 7027 and its central star}
\title{The Evolution of NGC 7027  at Radio Frequencies\\
A New Determination of the Distance and Core Mass}
\author{Albert~A.~Zijlstra}
\affil{Jodrell Bank Centre for Astrophysics,
Alan Turing Building, School of Physics and Astronomy,
The University of Manchester, Oxford Street,
Manchester M13 9PL, UK}
\email{a.zijlstra@manchester.ac.uk}
\and
\author{P.A.M.~van Hoof}
\affil{Royal Observatory of Belgium, Ringlaan 3, 1180 Brussels, Belgium}
\and
\author{R.A.~Perley}
\affil{National Radio Astronomy Observatory, P.O. Box O,
              Socorro, NM 87801, USA}
%\date{Accepted:
%      Received:
%      in original form }

\begin{abstract}
We present the results of a 25-year program to monitor the radio flux
evolution of the planetary nebula NGC\,7027. We find significant
evolution of the spectral flux densities.  The flux density at
1465~MHz, where the nebula is optically thick, is increasing at a rate
of $0.251\pm0.015\%\,\rm yr^{-1}$, caused by the expansion of the
ionized nebula. At frequencies where the emission is optically thin,
the spectral flux density is changing at a rate of $-0.145\pm0.005\%$
per year, caused by a decrease in the number of ionizing photons coming
from the central star. A distance of $980\pm100$\,pc is derived. By
fitting interpolated models of post-AGB evolution to the observed
changes, we find that over the 25-yr monitoring period, the stellar
temperature has increased by $3900\pm900$\,K and the stellar
bolometric luminosity has decreased by $1.75\pm0.38$\%.  We derive a
distance-independent stellar mass of $0.655\pm0.01\,\rm M_\odot$
adopting the Bl\"ocker stellar evolution models, or about
0.04\,M$_\odot$ higher when using models of Vassiliadis \&\ Wood which
may provide a better fit. A Cloudy photoionization model is used to
fit all epochs at all frequencies simultaneously. The differences
between the radio flux density predictions and the observed values
show some time-independent residuals of typically 1\%. A possible
explanation is inaccuracies in the radio flux scale of Baars {\sl et
al.}  We propose an adjustment to the flux density scale of the
primary radio flux calibrator 3C\,286, based on the Cloudy model of
NGC\,7027. We also calculate precise flux densities for NGC\,7027 for
all standard continuum bands used at the VLA, as well as for some new
30\,GHz experiments.
\end{abstract}


\section{Introduction}

Planetary nebulae trace one of the fastest phases of stellar
evolution. They form during the transformation from a low to
intermediate mass star (1--8\,M$_\odot$) to its final white dwarf
stage. The star first ejects its hydrogen envelope during a phase of
catastrophic mass loss.  This mass loss ceases once the hydrogen
envelope is reduced to $\sim 0.02\,\rm M_\odot$. The star subsequently
heats up, briefly reaching surface temperatures in excess of
$10^5$\,K, before the cessation of hydrogen burning when the remnant
C/O core enters the white dwarf cooling track. The evolution from the
end of the AGB mass loss to the onset of cooling lasts between $10^3$
and $10^5$\,yr: the duration is a steep function of white dwarf
mass \citep{Bloecker1995}. The time scales are short enough that one
may expect, in some cases, to observe significant changes in stellar
temperature and luminosity within decades.

Once the star has reached temperatures above $2 \times 10^4\,$K, it
ionizes the ejecta forming a bright planetary nebula (PN). The strong
emission lines and radio continuum from the PN provides a tracer of
the evolution of the parent star.  \citet{Masson1986, Masson1989} detected
evidence for expansion of the brightest planetary nebula, NGC\,7027,
using radio images over a four-year baseline: this provided proof of
the fast evolution. As the expansion velocity of the nebular gas is
known, the result provides both a measure of its distance and of its
age since ejection.  Masson found a distance of 880\,pc. The same
method has been applied by \citet{Hajian1993} who found a somewhat
faster angular expansion and therefore a lower distance, 770\,pc.  The
radio method has been applied to three other PNe \citep{Seaquist1991,
Christianto1998, Hajian1993, Kawamura1996}. Expansion-based distances
have also been measured using optical HST images \citep{Palen2002}.

The evolution of the temperature and luminosity of the star can be
derived through observations at optical or radio wavelengths. In the
optical, the absolute photometric accuracy is set by the Earth's
atmosphere as well as instrumentational limitations. At radio
frequencies, the required accuracies are currently achievable in the
centimeter wavelength range, due to the limited to negligible effect
of the atmosphere, the stability of radio telescopes and electronics,
and the availability of constant flux calibrators.  In this paper, we
utilize these to investigate variations in the the radio flux of
NGC\,7027 over a 25-year period.

NGC\,7027 is one of the secondary flux calibrators used in the
\citet{Baars1977} flux scale. Its rapid evolution may raise doubts on
its suitability as a calibrator.  However, the radio spectrum, due to
free-free emission, can be well understood and modeled, and hence
predicted.  We use this to investigate whether NGC\,7027 can be used to
improve the internal consistency of the radio flux calibration scale.

\section{Observations}

NGC\,7027 has been observed at the VLA over a 25-year period.  The
data analyzed are a subset of those taken as part of the VLA
Calibrator Flux Density project \citep{Perley2008}.  This
project commenced in 1983, with the goal of better determining the
flux density ratios between the secondary flux density calibrators
proposed in \citet{Baars1977}.  The VLA cannot by itself determine the
absolute flux density of any object, but can very accurately measure
the ratios between any pair of unresolved, or slighly resolved,
sources (at some frequencies, errors in the ratio are as low as
0.1\%).  These ratios can then be used to determine absolute spectral
flux densities, if at least one of the sources has a known spectral
flux density.  This can be provided by, for example, observations with
a telescope of known gain \citep[as reported by][]{Baars1977}, or by
observations of an object whose emission processes are sufficiently
well understood to permit calculation of its spectral flux density.

Observations in this program are typically made over the course of a
full day, once every $\sim16$ months when the VLA is in its
low-resolution `D'-configuration.  Short observations of each of 14
sources are taken at each available frequency band, approximately
hourly when the sources are between an elevation of 30 and 70 degrees.
These limits are set because of degraded high-frequency antenna
performance at low elevations (antenna gain loss due to gravitational
dish deformation, and increasing system temperatures due to
atmospheric emission) and because of degraded pointing performance for
observations near the zenith.  In 1983, the observations were made
only at the four original wavelength bands (20\,cm, 6\,cm, 2\,cm, and
1.3\,cm).  More recent observations have included the newer bands at
90\,cm, 3.6\,cm, and 0.7\,cm.  All epochs have data from five key
small-diameter steep-spectrum objects: 3C\,48, 3C\,138, 3C\,147, 3C\,286 and
3C\,295.  The first four of these are unresolved at all frequencies (in
the `D'-configuration), and form the basis of the antenna gain
calibration.  3C\,295 is known to be non-varying on physical grounds,
and hence establishes the reference for determining source
variability.

The dominant source of error for determination of accurate flux
densities at the shorter wavelengths (2\,cm, and less) is in antenna
pointing.  For all epochs after 1990, the technique of 'referenced
pointing' was utilized at the shorter wavelengths to minimize this
error.  In this technique, a nearby calibrator of known position is
utilized to determine the 'local' pointing parameters, which were then
applied to the target source.  This method is effective if the nearby
calibrator is within a few degrees. Unfortunately, for NGC\,7027, the
offset is rather large (eight degrees), limiting the effectiveness of
the correction, as we discuss later.  For the four basic calibrator
sources, the effectiveness is certain, as they served as their own
pointing calibrators.  Unfortunately, NGC\,7027 is too heavily
resolved to serve as its own referenced pointing source.

All data from all epochs were edited and calibrated following the same
procedures, summarized below.  These will be described at length in
\citet{Perley2008}.  

1) Clearly discrepant individual data were edited visually through
examination of visibility amplitudes.

2) Estimates of atmospheric opacity were made through `sky dips' at 23
and 43 GHz, and the visibility amplitudes corrected for absorption at
all frequencies, utlizing a simple atmospheric opacity model.  These
corrections are generally small, with zenith opacities typically 0.12
at 1.3\,cm, and 0.08 at 0.7\,cm.  

3) The antenna gain dependency on elevation was estimated by fitting a
polynomial to the observed elevation dependence of visibility
amplitudes (following the opacity correction) for the four unresolved
calibrators.  Residual errors in opacity will be absorbed into the
polynomial gain curve.  The ratios of the spectral flux densities of
the four sources are also determined at this step.

4) The \citet{Baars1977} value for the absolute flux density of 3C\,286
was assumed correct, and the flux densities of the remaining three
calibrator sources were determined after having accounted for opacity
and antenna gain dependencies.

5) The visibility amplitudes for all sources were then calibrated
utilizing the elevation-dependent gains and opacities, and the
calibrator source flux densities, through application of a temporally
constant gain solution.

6) The data were then re-examined for discrepant values -- most
notably due to antenna pointing errors -- and the preceding steps were
repeated until the residuals in the derived antenna gains showed an
absence of clearly deviant values.

3C\,286 is utilized as the flux density standard as it is unresolved,
and its flux density ratio to 3C\,295 has been determined to be changing
by less than 0.01\%/year at all bands \citep{Perley2008}.  

Removal of atmospheric phase perturbations was done through the
well-established methods of self-calibration, whereby the source
emission is used to estimate and correct the antenna-based phase
errors, utilizing a rough initial estimate of the source structure.
In general, the process is iterative, as the source structure is not
known in advance.  However, for NGC\,7027, we have utilized all the
data, over all epochs, to establish a standard model to accelerate the
convergence.  Only the phases are corrected in this process, while the
amplitudes are held fixed, to prevent `wandering' of the flux
density.  At the resolutions utilized in this program, the secular
changes in angular size are completely negligible, thus allowing the
use of a single model.  

Following basic calibration, the spectral flux density for each target
source at each frequency and epoch was then determined by a number of
methods, depending on source size and observing frequency.  For NGC
7027 at 20\,cm, the large primary beam includes many background
sources whose contributions must be removed from that of NGC\,7027:
these background sources contribute $\sim$265 mJy of additional flux
density to the $\sim$1.5\,Jy from NGC\,7027.  The separation is done
through imaging by fourier inversion.  All data from the observations
for each epoch were gridded, fourier transformed, and deconvolved
following established methods.  The effective resolution of 45
arcseconds partially resolves the source, so the integrated flux
density is derived by spatial integration of the imaged response.  Two
methods were utilized: one simply sums all the brightness within the
observed bounds of the source, while the other fits a gaussian to the
observed brightness.  This latter method can only be used for very
slightly resolved objects, an assumption well justified at this
frequency.

At all other frequencies, NGC\,7027 is heavily resolved, while
background source contamination is negligible as the antenna primary
beam restricts the field of view.  For these, the source emission was
imaged, and the total flux estimated by spatial integration over the
source.

The uncertainty in the flux density determination for a particular
epoch and frequency was determined by measuring the source flux for
each of the $N \sim$\,8 to 10 observations taken at each band for each
epoch.  The variance in these determination, divided by $\sqrt{N-1}$,
gives a 1-$\sigma$ estimate of the error.  The justification that the
$N$ observations are independent arises from analysis of pointing
errors from high-frequency observations of standard calibrators: the
typical angular scale for pointing variations (due to the antenna
azimuth and elevation mounts) is found to be $\sim$15 degrees, or
about 1 hour in time.  At longer wavelengths (3.6, 6 and 20\,cm),
pointing errors are not the dominant source of error, which is likely
to be set by small-temporal-scale variations in antenna gain.  The
timescale for these is not certain, but is likely to be less than an
hour.

\section{NGC\,7027 radio flux densities}

Data were obtained for 12 epochs between 1983 and 2006.  The available
data, along with the estimated uncertainties, are summarized in Table
\ref{ratios}, as ratios between NGC\,7027 and 3C\,286. Flux densities
for NGC\,7027 can be calculated using the Baars flux density scale for
3C\,286, listed in Table \ref{flux_scale}. The Baars scale for 3C\,286
extends only to 15 GHz -- for the 23 and 43 GHz calibration, we have
utilized observations of, and a model for the planet Mars, as will be
described by \citet{Perley2008}. In this paper we will refer to the
full frequency range as the 'Baars scale', even where the supplemental
Mars data is used. The Baars et al. 3C\,286 values are likely to have their
own uncertainties, as we argue for below.

\vspace*{2cm}
\begin{deluxetable}{lllllllllllllllllll}
\tabletypesize{\scriptsize}
\tablewidth{20.8cm}
\rotate
\tablecaption{\label{ratios} NGC~7027/3C\,286 flux density ratios, with uncertainties.}
\tablehead{
Epoch & \multicolumn{2}{c}{1275\,MHz}  & \multicolumn{2}{c}{1465\,MHz}  &        
\multicolumn{2}{c}{4535\,MHz}  &  \multicolumn{2}{c}{4885\,MHz}  &       
\multicolumn{2}{c}{8435\,MHz} &   \multicolumn{2}{c}{8735\,MHz} &
\multicolumn{2}{c}{14965\,MHz}  &  \multicolumn{2}{c}{22460\,MHz}  &
\multicolumn{2}{c}{43340\,MHz} \\
 & ratio & unc. & ratio & unc. & ratio & unc. & ratio & unc. & ratio & unc. & ratio & unc.  & ratio & unc. & ratio & unc. & ratio & unc.
}
\startdata
1983.42	& \nodata& \nodata &  .1003  & .0004 & \nodata& \nodata&  .7413 & .0022
 & \nodata& \nodata& \nodata&  \nodata	& 1.67   &  .06   & 2.29  & .13   & \nodata & \nodata  \\
1985.99	& \nodata&  \nodata &  .1014  & .0002 & \nodata& \nodata&  .7457 & .0014 
 & \nodata& \nodata&  \nodata & \nodata & 1.716  & .006   & 2.206 & .053  & \nodata & \nodata \\
1987.34	& \nodata& \nodata &  .1022  & .0004 & \nodata& \nodata&  .7498 & .0016 
 & 1.183 & .004  & \nodata & \nodata & 1.760  & .016   & 2.292 & .044  & \nodata & \nodata \\
1989.99	& \nodata& \nodata &  .1022  & .0003 & \nodata& \nodata&  .7424 & .0015 
 & 1.166 & .004  & \nodata & \nodata & 1.715  & .011   & 2.222 & .019  & \nodata & \nodata\\
1995.20	& \nodata& \nodata &  .1043  & .0005 & \nodata& \nodata&  .7411 & .0022 
 & 1.165 & .002  & \nodata & \nodata & 1.7013 & .0028  & 2.183 & .007  & 3.21    & .12\\
1998.09	& \nodata& \nodata &  .1049  & .0004 & \nodata& \nodata&  .7374 & .0015 
 & 1.159 & .0018 & \nodata & \nodata & 1.692  & .003   & 2.214 & .021  & 3.15    & .10\\
1999.35	& .0758  & .00026  &  .10467 & .00010 & \nodata& \nodata&  .7400 & .0009  
 & 1.153 & .0011 & \nodata & \nodata & 1.689  & .003   & 2.199 & .008  & 3.27    & .04\\
2000.77	& .07494 & .00060 &  .10481 & .00068 &  .6860 & .0008 &  .7396 & .0010
 &  1.150 & .003   & 1.18   & .0017  & 1.693  &  .004  & 2.154 & .009  & 3.20    & .06 \\
2001.86	& .0770 &  .0006  &  .1063  & .0004 &   .6830 & .0011 &  .7356 & .0015
 & 1.155 & .0019 &  1.1844 & .002   & 1.686  & .008   & 2.15  & .017  & 3.29    & .06\\
2003.10	& .07673&  .0002  &  .1060  & .0002 &   .6833 & .0010 &  .7362 & .0010
 & 1.155 & .0020 &  1.186  & .002  & 1.682  & .005   & 2.199 & .008  & 3.22    & .040 \\
2004.65	& .0763 &  .00028 &  .1066  & .00069 &  .6835 & .0013 &  .7352 & .0009
 &  1.150 & .0010 &  1.1768 & .0013 & 1.6824 & .0028  & 2.199 & .013  & 3.368   & .040 \\
2006.04	& .0774 &  .00033 &  .10668 & .00029 &  .6822 & .0013 &  .7350 & .0014
 &  1.145 &  .0020 & 1.182  & .0020 & 1.674  &  .006  & 2.119 & .010  & 3.07    & .040 \\
\enddata
\end{deluxetable}

\begin{table*}
\caption[]{\label{flux_scale} The spectral flux density of 3C\,286.
 The values in the first flux density row are referred to as the
'Baars scale', where the 23 and 43\,GHz values come from a model for
Mars, as the original Baars scale does not extend to these
frequencies.  The last row gives 'corrected' fluxes, derived using the
NGC\,7027 model described in this paper.  }
\begin{flushleft}
\begin{tabular}{llllllllll}
\tableline
\tableline
Freq. [MHz]       & 1275  &  1465  &  4535  &  4885  & 8435
                  & 8735  &  14965 &  22460 & 43340 \\
\tableline
Flux density [Jy] & 15.52 & 14.51  &  7.758 &  7.410 &  5.189
                  & 5.066 &  3.455 &  2.560 & 1.537 \\
                  & 15.33 & 14.70  &  7.875 & 7.477  & 5.132
                  & 5.004 & 3.439  &  2.571 & \nodata\\
\tableline
\end{tabular}
\end{flushleft}
\end{table*}

At the higher frequencies, NGC\,7027 shows a larger scatter in the
radio flux, especially during the first two epochs, than at the lower
frequencies.  At these earlier times, the VLA pointing was not as well
established, nor were the receivers as sensitive as those now
employed.  In addition, referenced pointing was not employed until
1990.  

\begin{figure*}
    \hbox{\includegraphics[width=65mm]{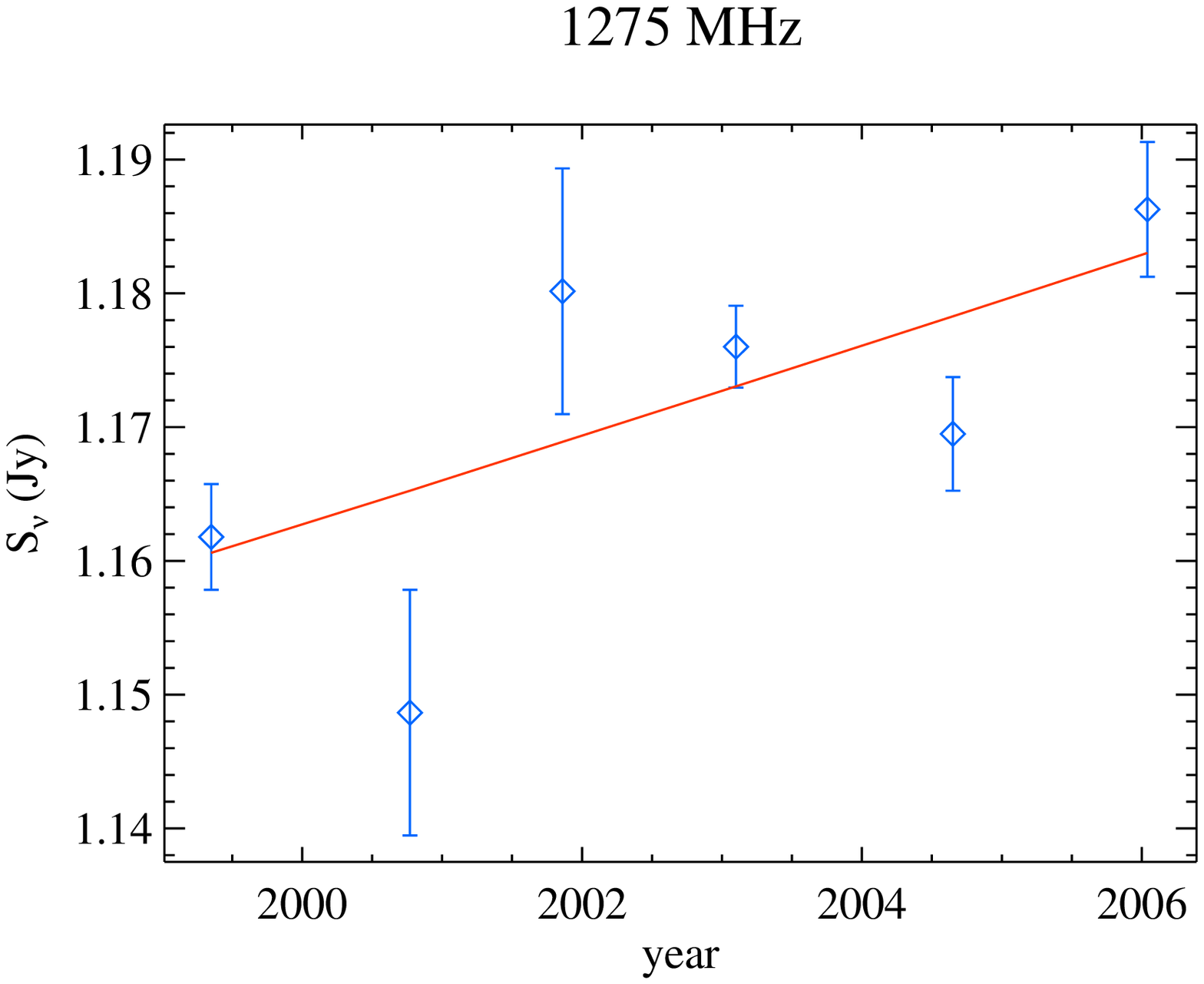}
          \hspace{5mm}
          \includegraphics[width=65mm]{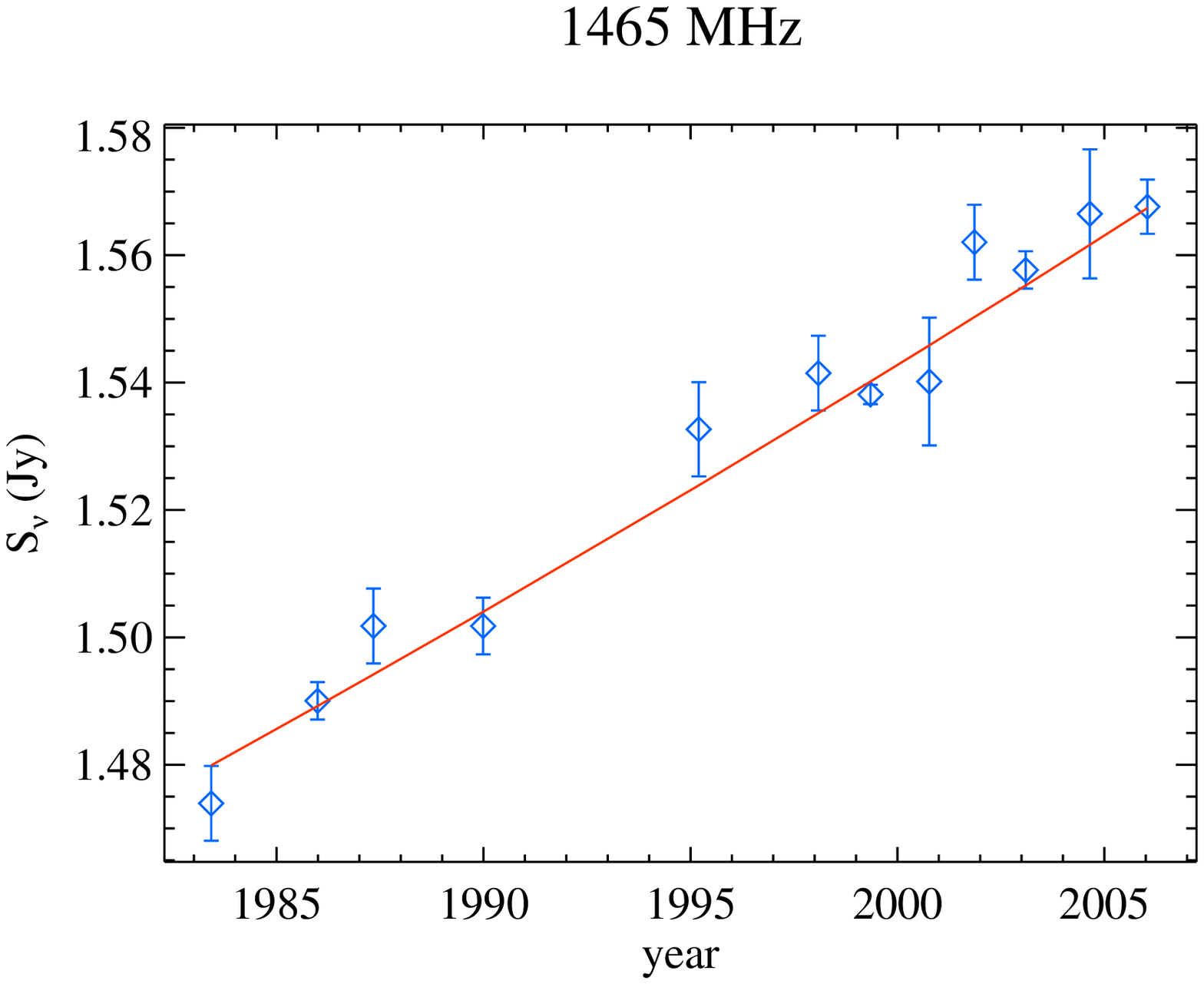}
    }
    \vspace{3mm}
    \hbox{\includegraphics[width=65mm]{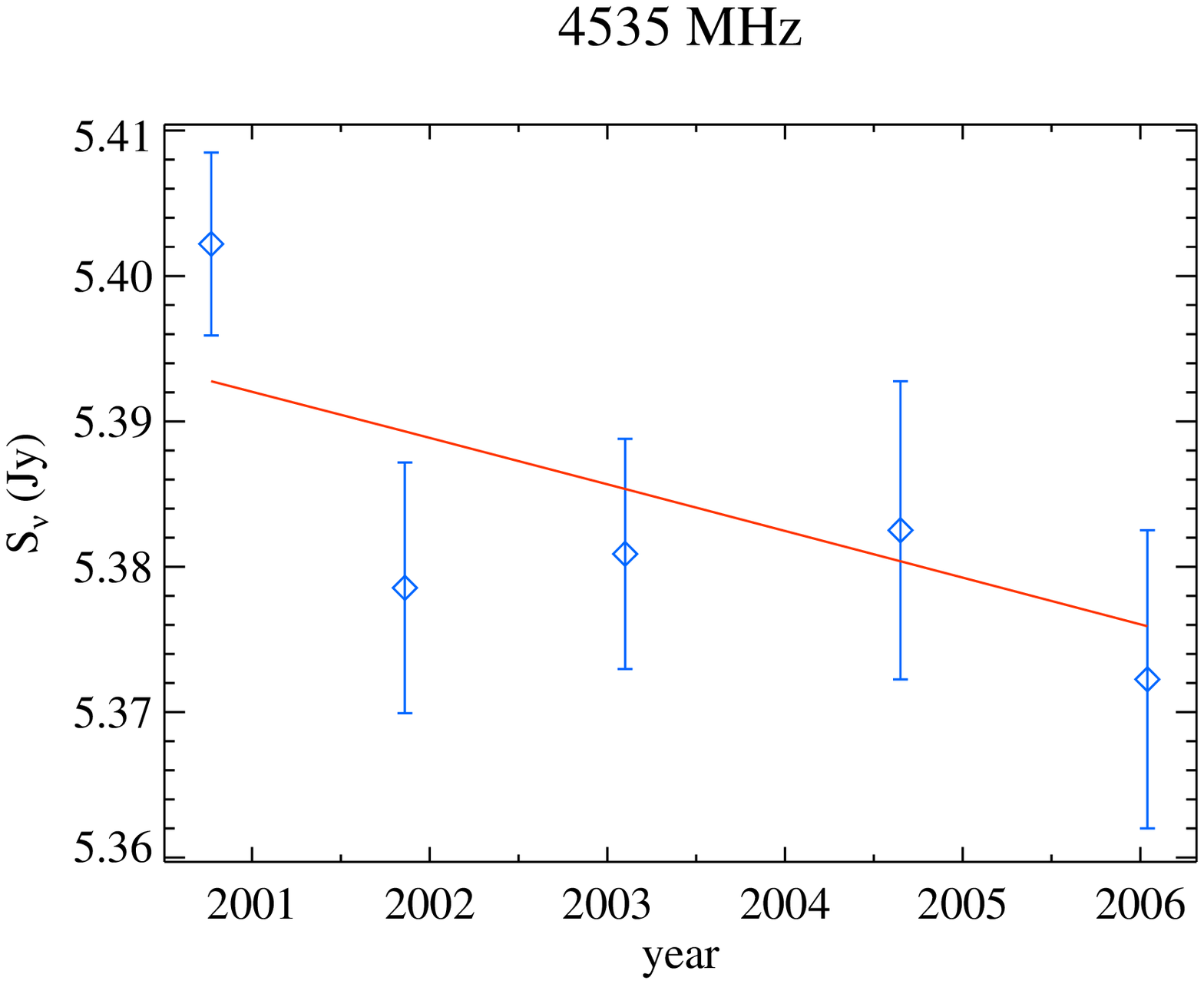}
          \hspace{5mm}
          \includegraphics[width=65mm]{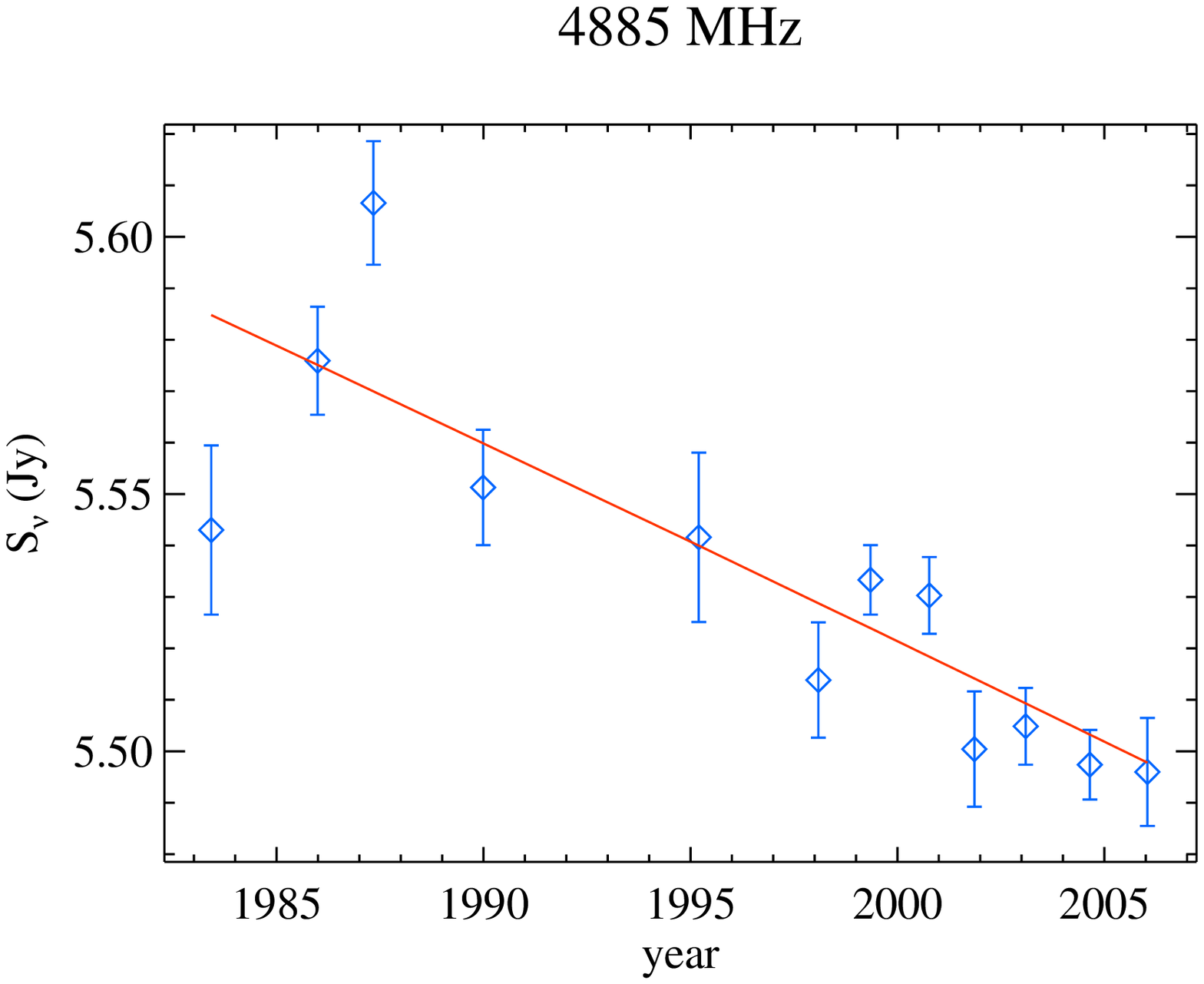}
    }
    \vspace{3mm}
    \hbox{\includegraphics[width=65mm]{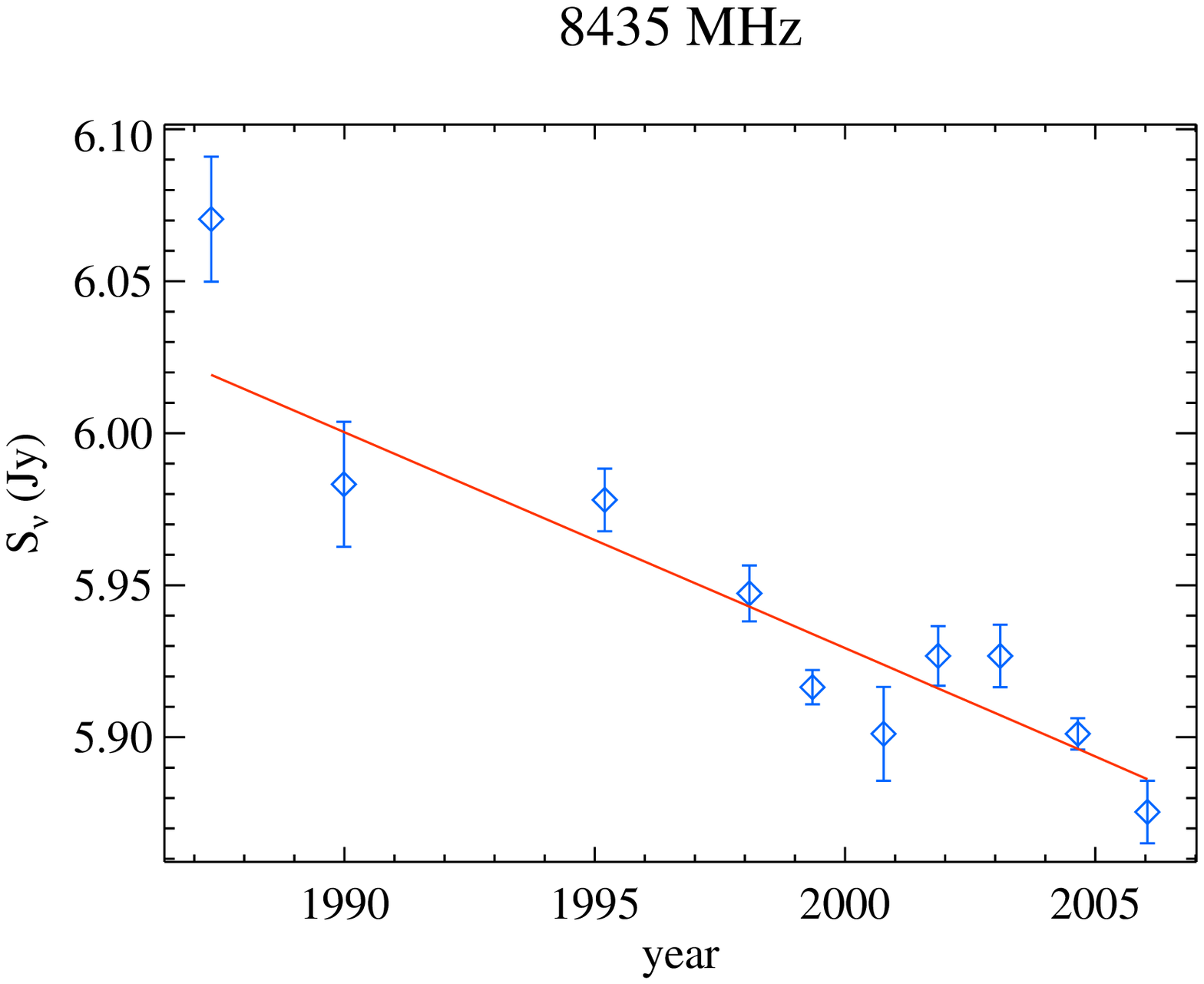}
          \hspace{5mm}
          \includegraphics[width=65mm]{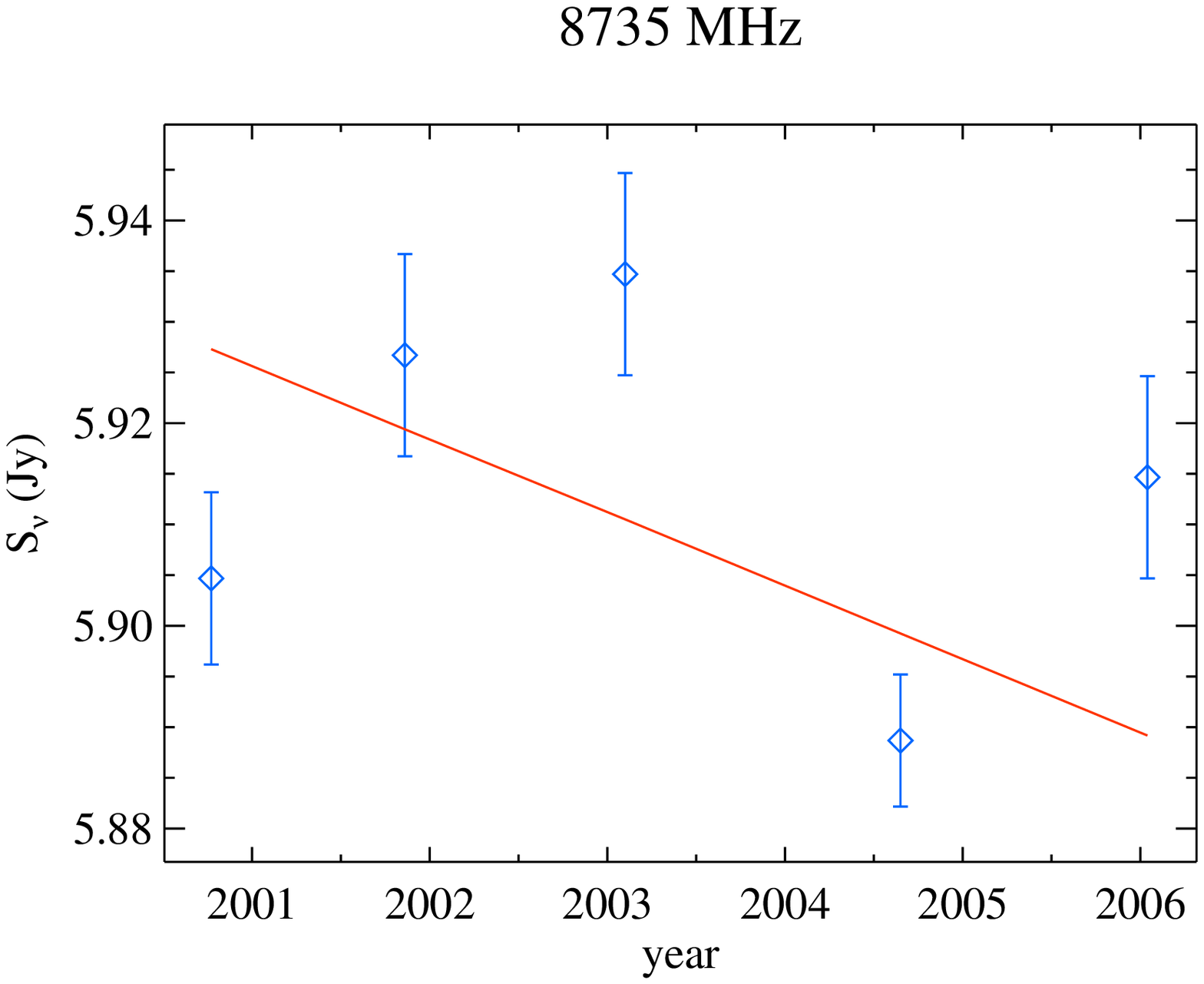}
    }
    \vspace{3mm}
    \hbox{\includegraphics[width=65mm]{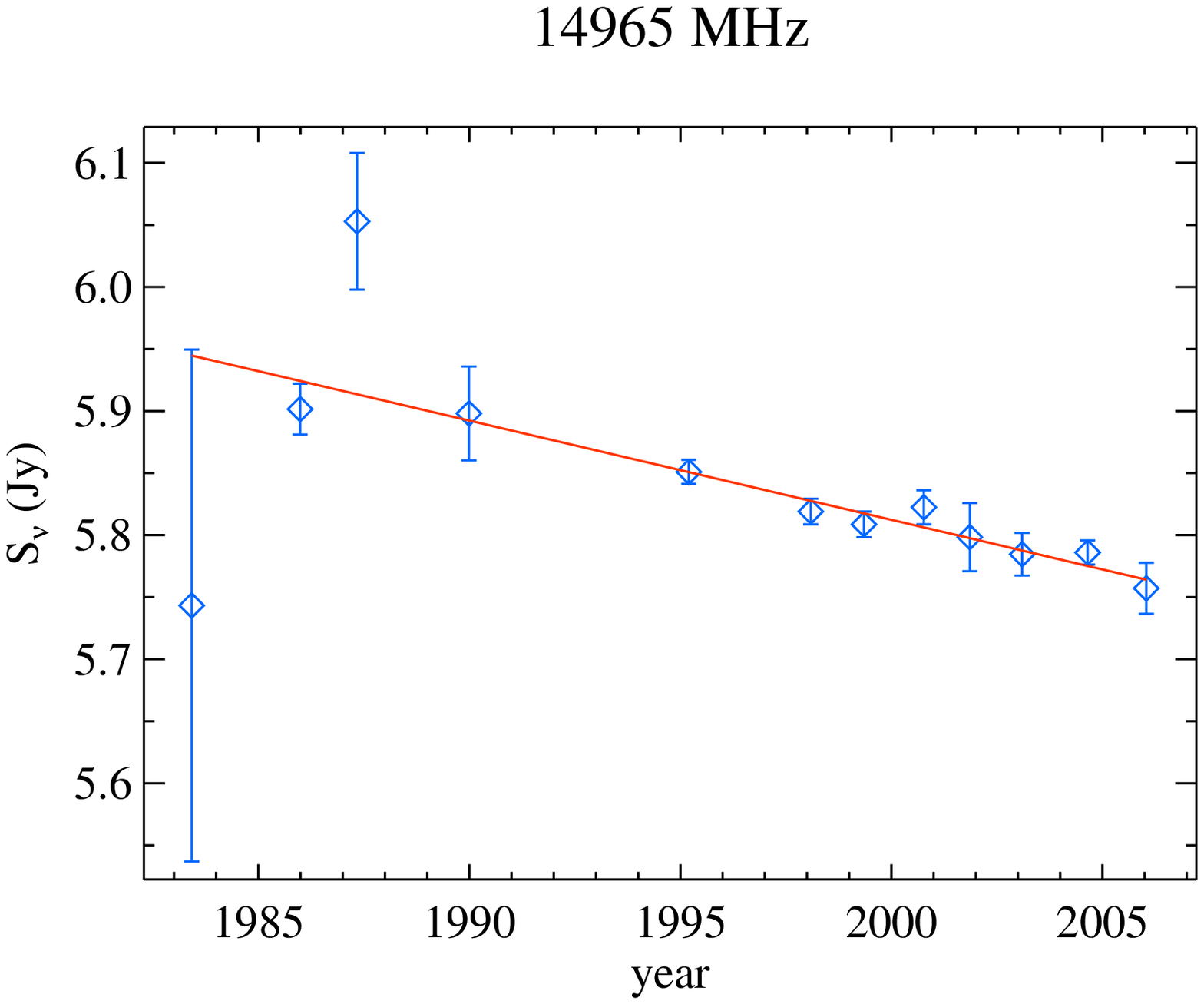}
          \hspace{5mm}
          \includegraphics[width=65mm]{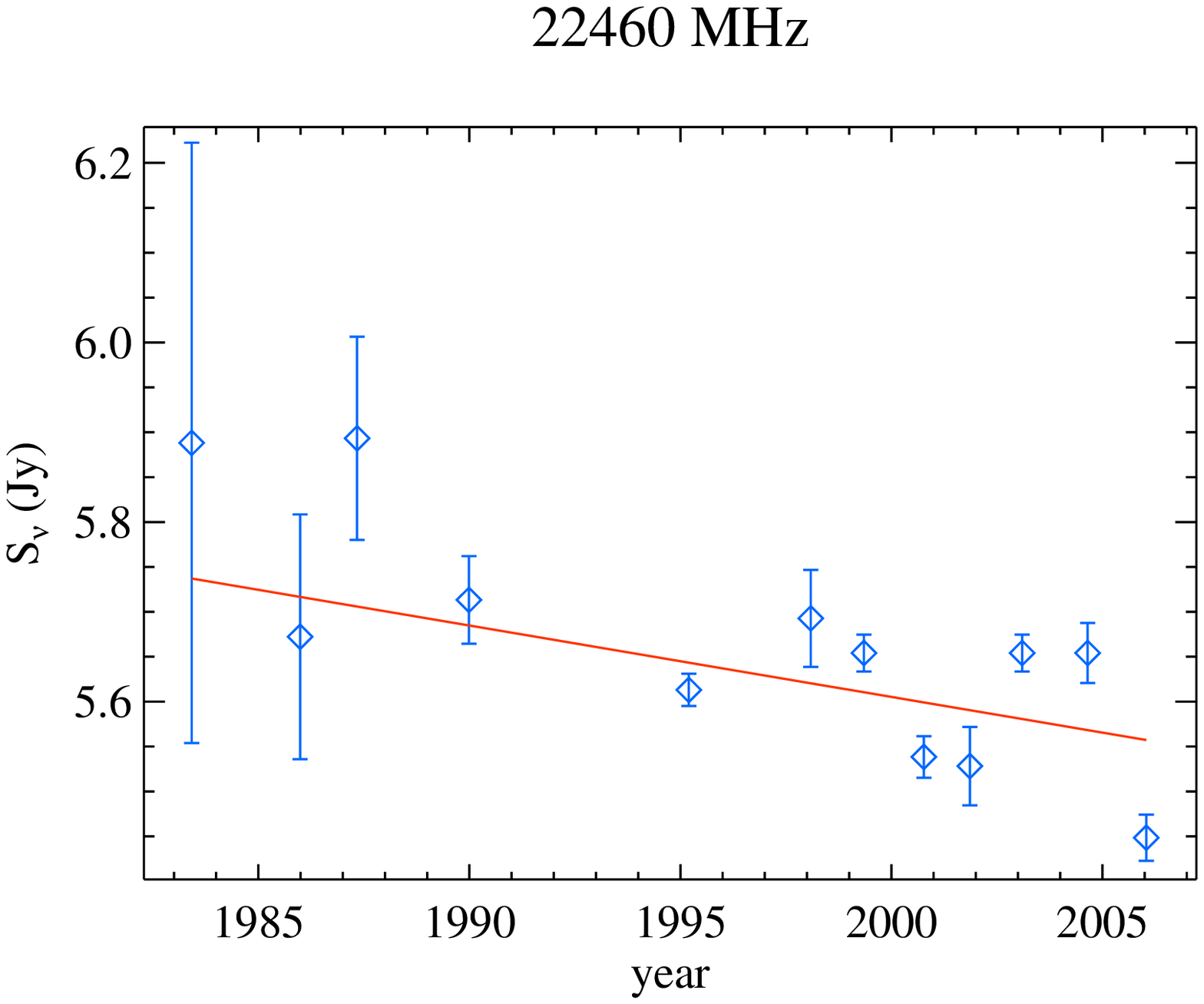}
    }
    \caption{\label{changes}NGC 7027 radio spectral flux evolution. 
    The drawn line shows the fit from the photo-ionization
 models, discussed in the text. The observed fluxes are normalized on
 the revised Baars scale, as discussed in Sect.~\ref{baars_revised}.}
\end{figure*}

The 43\,GHz fluxes show large fluctuations, and are also considerably
below the expected flux density for this frequency, compared, e.g., to
the 1998 measurement of the absolute flux at 32\,GHz, of
$5.5\pm0.2\,$Jy \citep{Mason1999}. The uncertainties may be
weather-related, as the single epoch (2004.65) where the flux
approached its expected value had exceptionally good (and dry)
weather, while the 2006 observations with extremely low flux had very
poor (and windy) weather.  It is also probable that residual pointing
errors are responsible -- pointing errors can only lower the observed
flux, and the source is unique amongst all the sources in the program
in that it both transits near the zenith (where VLA pointing is most
problematic) and that the nearest referenced pointing calibrator is
over eight degrees away -- a separation amplified by $1/\cos(E)$ in
the antenna's coordinate frame, where $E$ is the elevation.  These
problems are probably also responsible for the larger errors in the
22460 MHz observations, as shown in Figure \ref{changes}.

After allowing for the uncertainties, the ratios show significant,
slow variations with time, well above the uncertainties.  The changes
in the fluxes are plotted in Fig. \ref{changes}. At most frequencies,
evidence is seen for linear changes with time. The 43\,GHz data are not
shown, and have not been incorporated into the analysis, for the
reasons detailed above.

Interestingly, the secular changes vary with frequency. At 1.4\,GHz,
where the nebula is optically thick, a significant, steady flux
increase is seen, from 1.46 to 1.56\,Jy over the time span of the
observations. At all higher frequencies, where the nebula is partially
or fully optically thin, a small but significant decrease is seen
instead.  The fractional change per year ranges from +0.251 (at 1465~MHz)
to $-0.145$ per cent per year at optically thin frequencies.

\section{SED modeling}

The model of the radio spectrum is based on the {\sc Cloudy}
photoionization model described in \citet{Beintema1996}. This
one-dimensional model was used to fit the rich ISO infrared line
spectrum, as well as the radio, optical and UV spectrum; a stellar
(blackbody) temperature of $T_{\rm eff}=1.61 \times 10^5\,\rm K$ was
derived, with some evidence for changes after 1981 which could be due
to an increase in temperature.

We modified this Cloudy model by forcing the inner radius to 2/3 of
the outer radius, in order to give better agreement with the observed
thickness of the shell as reported by \citet{Masson1989}. A new fit
was derived, to give the best fit to all the observational data,
including the 15\,GHz radio flux measurement from
\citet{Masson1989}. This model yields the electron temperature and
density, as well as the ionization structure of the most important
elements (H, He, C, N, O) as a function of radius. The abundances
determined with the new Cloudy model showed some
differences. Especially the helium abundance changed from 10.98 in
\citet{Beintema1996} to 11.08 in the current model. These abundances
are relative number densities on a logarithmic scale w.r.t. n(H)
$\equiv 10^{12}$. Abundances have also been derived from this ISO
spectrum by \citet{Salas2001}, but using ionization correction
factors, rather than modeling. We tested the sensitivity of the
results discussed below to the abundance differences, but found only
insignificant changes. The new model requires a somewhat higher
stellar temperature to fit the high ionization lines, because of the
larger inner radius, $T_{\rm eff}=1.9 \times 10^5\,\rm K$. Because we
find that this is very sensitive to the adopted inner radius, below we
will use the temperature from \citet{Beintema1996}.

The radio fluxes were calculated using a purpose-built spherically
symmetric model. This included a very accurate Gaunt free-free factor
routine and exact radio radiative transfer to cover optical depth
effects.  Radio emission can originate from other processes than
free-free emission.  Non-thermal emission is not observed in PNe, but
rotational emission from dust has been suggested \citep{Casassus2004},
giving excess emission at 20-30\,GHz.  Sub-millimeter polarimetry has
confirmed the presence of spinning, non-spherical dust grains in
NGC\,7027 \citep{Sabin2007}. However, using emissivities tabulated
in \citet{Draine1998}, we find an expected flux from spinning dust at
22\,GHz of $\sim 1\,$mJy, which is negligible. Radio recombination lines
are also not included in the model. All frequency bands used avoid 
the H$n\alpha$ transitions, apart from the 4.885~GHz band which includes
the H110$\alpha$ line with a line-to-continuum ratio of $0.40\pm0.09$\%
\citep{Roelfsema1991}: this line will contribute $<0.1$\%\ to the total
in-band flux. 

Four free parameters were introduced to model secular variations in
the number of ionizing photons (due to evolution of the central star)
and optical depth scale (due to expansion of the nebula). Both were
assumed to vary linearly, of the form $a + b\times(t-2000)$ where the
scaling factors are independent of frequency, and $t$ is time in
years.

The electron temperature can be considered constant with time.  The
photoionization equilibrium is established on a time scale of the order
of a year. The cooling function of nebular gas is a very steep
function of electron temperature. Hence, obtaining an appreciable amount
of change in the electron temperature requires a large change in the
heating rate: the so-called thermostat effect.

\begin{table}
\caption[]{\label{modelflux} The uncorrected model fluxes for
NGC\,7027, valid for 2000.0, together with the secular changes.  The
last column gives a multiplicative factor for each flux, implied by the
full fit. }
\begin{flushleft}
\begin{tabular}{rccc}
\tableline
\tableline
     freq &   flux  & sec.var. & f \\
      GHz &    Jy   &  mJy/yr  &  \\
\tableline
    1.2750 & 1.1627 &  $+$3.28 & 1.0126 $\pm$ 0.0006 \\
    1.4650 & 1.5428 &  $+$4.00 & 0.9874 $\pm$ 0.0006 \\
    4.5350 & 5.3953 &  $-$3.18 & 0.9852 $\pm$ 0.0006 \\
    4.8850 & 5.5215 &  $-$3.88 & 0.9910 $\pm$ 0.0005 \\
    8.4350 & 5.9294 &  $-$7.12 & 1.0112 $\pm$ 0.0005 \\
    8.7350 & 5.9329 &  $-$7.23 & 1.0124 $\pm$ 0.0006 \\
   14.9650 & 5.8125 &  $-$7.99 & 1.0046 $\pm$ 0.0007 \\
   22.4600 & 5.6054 &  $-$7.95 & 0.9956 $\pm$ 0.0006 \\
   43.3400 & 5.2034 &  $-$7.51 & \nodata \\
\tableline
\end{tabular}
\end{flushleft}
\end{table}

Using this procedure, the quality of the fit does not agree with the
uncertainties on the data: the model flux densities are typically discrepant
by $\pm1$\% -- offsets significantly greater than the measurement
uncertainties. We therefore decided to include in the model systematic
uncertainties in the flux density scale, represented by multiplicative factors
for each frequency which are constant over time. We set the conditions that
the factors for 1275 and 1465 MHz add up to 2, and the factors for the higher
frequencies sum to 6 exactly, to avoid degeneracy with the scale factors '$a$'
defined above (the 43\,GHz flux densities were not used for reasons described
earlier). This division is based on the observation that the lowest two
frequencies will be mainly sensitive to the scale factor for the optical
depth, while the higher frequencies will be mainly sensitive to the scale
factor for the number of ionizing photons. In effect, we have by this
procedure fixed the flux density scale to the Baars scale including the Mars
calibration at 23 GHz. Observations at different frequencies will have
different weights, determined by the number of observations, the observational
uncertainty, and the strength of the optical depth effect. This procedure
resulted in a total of 10 free parameters for 74 observations.

The multiplicative factors gave a large improvement, with the $\chi^2$
reducing from 2000 to 170. The resulting best fit is shown in Table
\ref{modelflux}, where the flux values refer to the epoch 2000.0. One would
expect an ideal $\chi^2=64$. The larger $\chi^2$ obtained may indicate that
the errors on the flux determination have been underestimated, especially at
23\,GHz.

The best fit is given by the following relations, where $Q({\rm H})$
is the number of hydrogen-ionizing photons emitted per second,
$\tau_\nu$ is the optical depth at radio frequencies and $t$ is the
time in years.

\begin{eqnarray}
\nonumber
 Q({\rm H}) &=& Q({\rm H})[2000] \times ( 1 + c \times (t - 2000) ), \\
 \tau_\nu &=& \tau_\nu[2000] \times (1 + d \times (t - 2000) ), 
\end{eqnarray}
with
\begin{eqnarray}
%\nonumber
%    a &=&  1.00000 \pm 0.00023 \\
\nonumber
    c &=& -0.00145 \pm 0.00005, \\
%\nonumber
%    c &=&  1.00000 \pm 0.00059 \\
    d &=& -0.00410 \pm 0.00010. 
\end{eqnarray}

The secular variation of the model flux is in principle
non-linear, but this is only noticeable (barely) for the lowest frequencies.

\section{Expansion and distance}

\subsection{Measured expansion}

The flux density in the optically thick regime, at 1465.9\,MHz, is increasing
at a rate of $0.251\pm0.015$ per cent\,yr$^{-1}$. We note that the brightness
temperature, which is given by

\begin{equation}
  T_{\rm b} = T_{\rm e} \left(1 - e^{-\tau}  \right),
\end{equation}

\noindent derived from high-resolution 1.4\,GHz data \citep[$T_b=15$--$17\,\rm
kK$:][]{Bains2003} is in good agreement with the electron temperature
predicted by the Cloudy model ($T_e=15\,\rm kK$). This indicates that
the optical depth at these frequencies must be high. The intensity at
the surface of the nebula is approximately equal to the source
function $S_\nu$ at an optical depth of 2/3. Since the source function
can be approximated by the blackbody function ($S_\nu \approx
B_\nu(T_{\rm e})$) in the radio regime, it follows that the intensity
only depends on the electron temperature, which can be assumed to be
nearly constant, as we already argued above. Hence the flux density
only depends on the surface area of the emitting region. Since the
total optical depth in the nebula is high, the layer at which $\tau =
2/3$ is reached must be close to the ionization front. Therefore the
increase in the 1465.9\,MHz is measures the increase in the surface
area of the ionized region.

The uncertainty on the flux increase is approximately 6\%. This is a
factor of two better than that of the direct (1-d) expansion
measurements of \citet{Hajian1993} and \citet{Masson1989}.  This
warrants a new derivation of the distance, where we derive the expansion
through the flux increase.

\subsection{Distance}

\begin{figure}
 \hbox{\includegraphics[width=65mm,clip=]{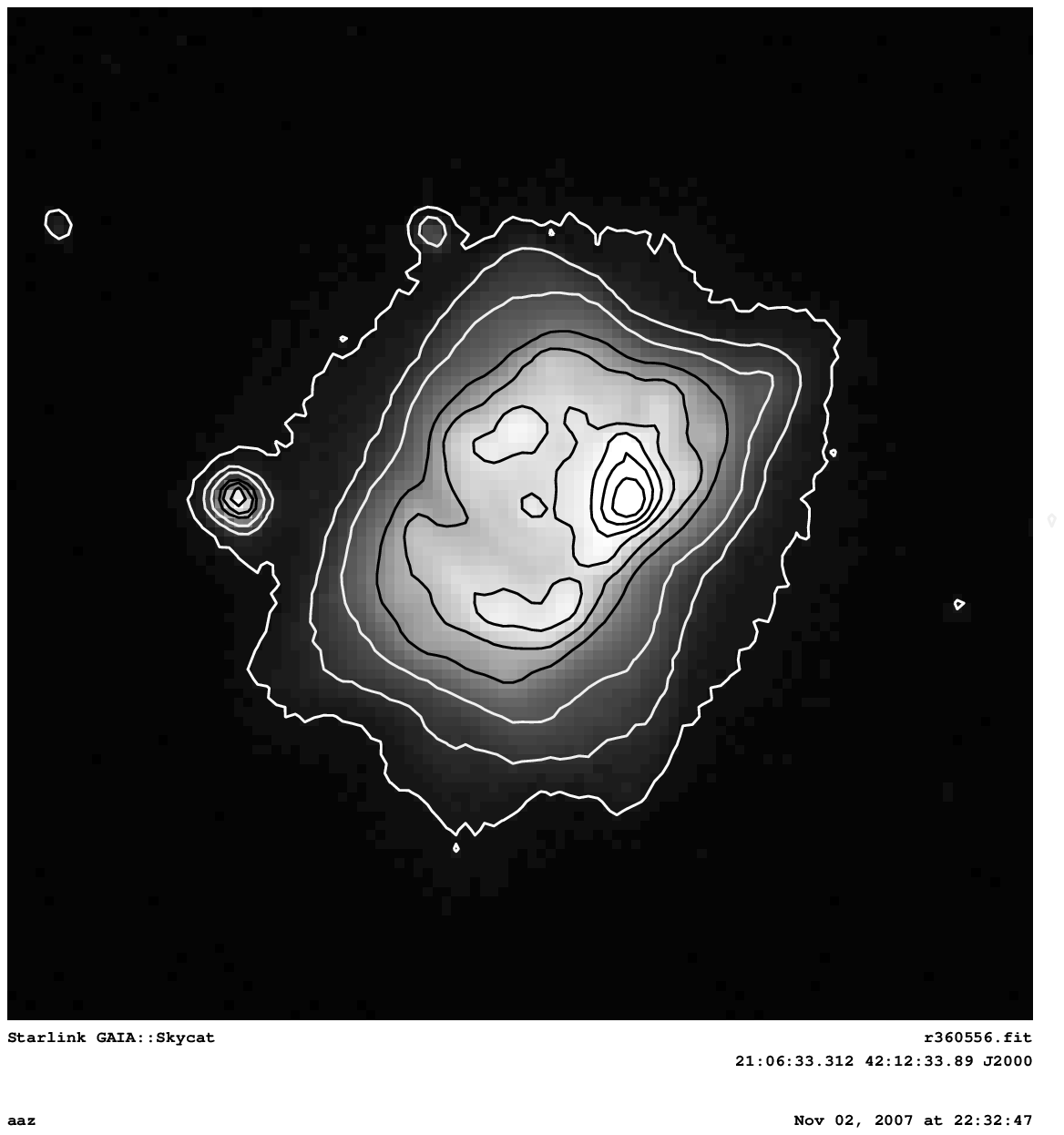} \hspace{5mm}
  \includegraphics[width=70mm,clip=]{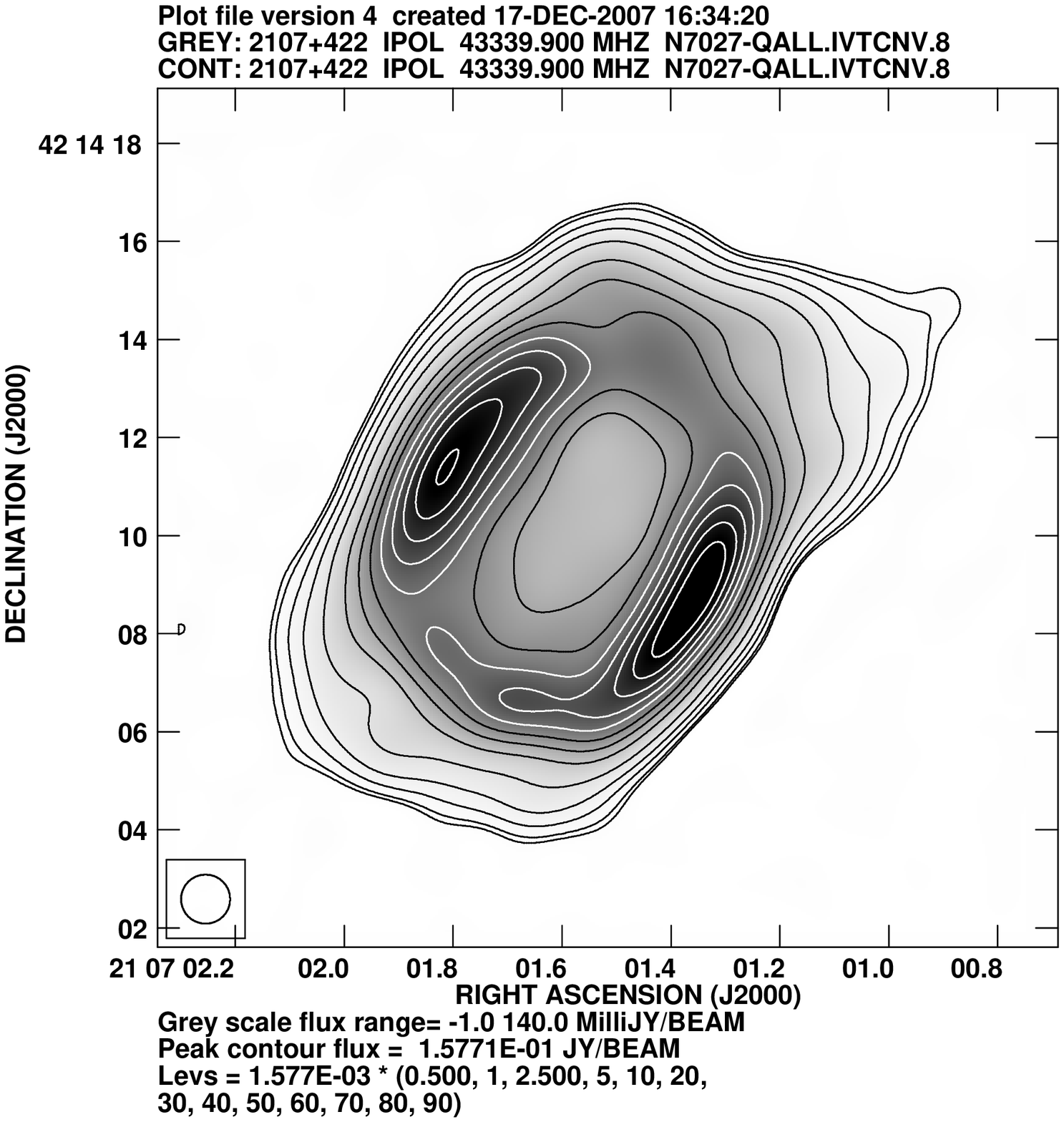}
  }\caption{\label{iphas}Left: An i-band image of NGC\,7027, taken
  from the IPHAS survey. The frame is approximately 35 arcsec across;
  north is up and east is left.  Right: a VLA image with 1''
  resolution, obtained from the 43GHz data. The inset shows the size
  of the beam}
\end{figure}

To obtain a distance from the measured angular expansion, an expansion
velocity is needed.  The expansion velocity in the equatorial plane is
reported as $v_{\rm exp}=13\pm1 \,\rm km\,s^{-1}$, based on the [O\,{\sc
iii}] line \citep{Bains2003}. This value was used by these authors,
together with the measured angular expansion \citep{Masson1989} to
obtain a rather short distance of $650\pm100$\,pc.  Other measured
values of the velocities are higher, such as the
19.5\,km\,s$^{-1}$ in Br$\gamma$ \citep{Cox2002}. In optical lines 
\citet{Walsh1997} find an expansion velocity  of
$v_{\rm minor} = 18.9\rm \,km\,s^{-1}$.  The CO shell has an expansion
velocity of 13--16\,km\,s$^{-1}$ \citep{Hasegawa2001, Fong2006}.  Other lines
arising from the photo-dissociation region show larger line widths
than does the CO.  High velocity gas, reaching 55\,km\,s$^{-1}$, is
seen along a bipolar axis \citep{Cox2002}, but such a fast component
is not known from the ionized gas. Radio recombination lines indicate
an expansion velocity of $21\pm2\,\rm km\, s^{-1}$ \citep{Ershov1989}.

Most of the measurements refer to the equatorial direction where most
of the gas is located, and where the 1-d radio expansion was
measured. Our 2-d expansion measurement also requires information on
the polar expansion, which is known to be faster.  The velocity field
has been studied comprehensively by \citet{Walsh1997}
and \citet{Roelfsema1991}. The velocity law obtained by fitting the
3-dimensional line data (optical forbidden lines and radio
recombination lines) is \citep{Walsh1997}:

\begin{equation}
\label{vfield}
  v = -10 + 1400 \frac{r \rm[pc]}{D \rm[kpc]}
\end{equation}

\noindent where $r$ is the radial coordinate in the nebula and $D$ is
the distance. The velocities are radially away from the central star. This
model predicts a peak velocity on the minor axis of $v_{\rm minor}=
14.5\rm \,km\,s^{-1}$, and on the major axis $v_{\rm major}= 29\rm 
\,km\,s^{-1}$. 

The velocity law of \citet{Walsh1997} does not preserve the
morphology: the  expansion is not self-similar. This is  expected,
as for bipolar planetary nebulae the opening angle of the polar lobes 
increases over time, eventually reaching an 'X' shape. NGC\,7027 is still
at an early phase of this evolution.

The ionized nebula is approximately rectangular with diameter
$7.2\times11.5$ arcsec, measured from the 1.4\,GHz images presented
in \citet{Bains2003}. An $i$-band optical image obtained during the
IPHAS survey \citep{Drew2005}, and a VLA image derived from the
43\,GHz data, are shownn in Fig. \ref{iphas}. 

\begin{figure*}
\includegraphics[width=170mm, clip=]{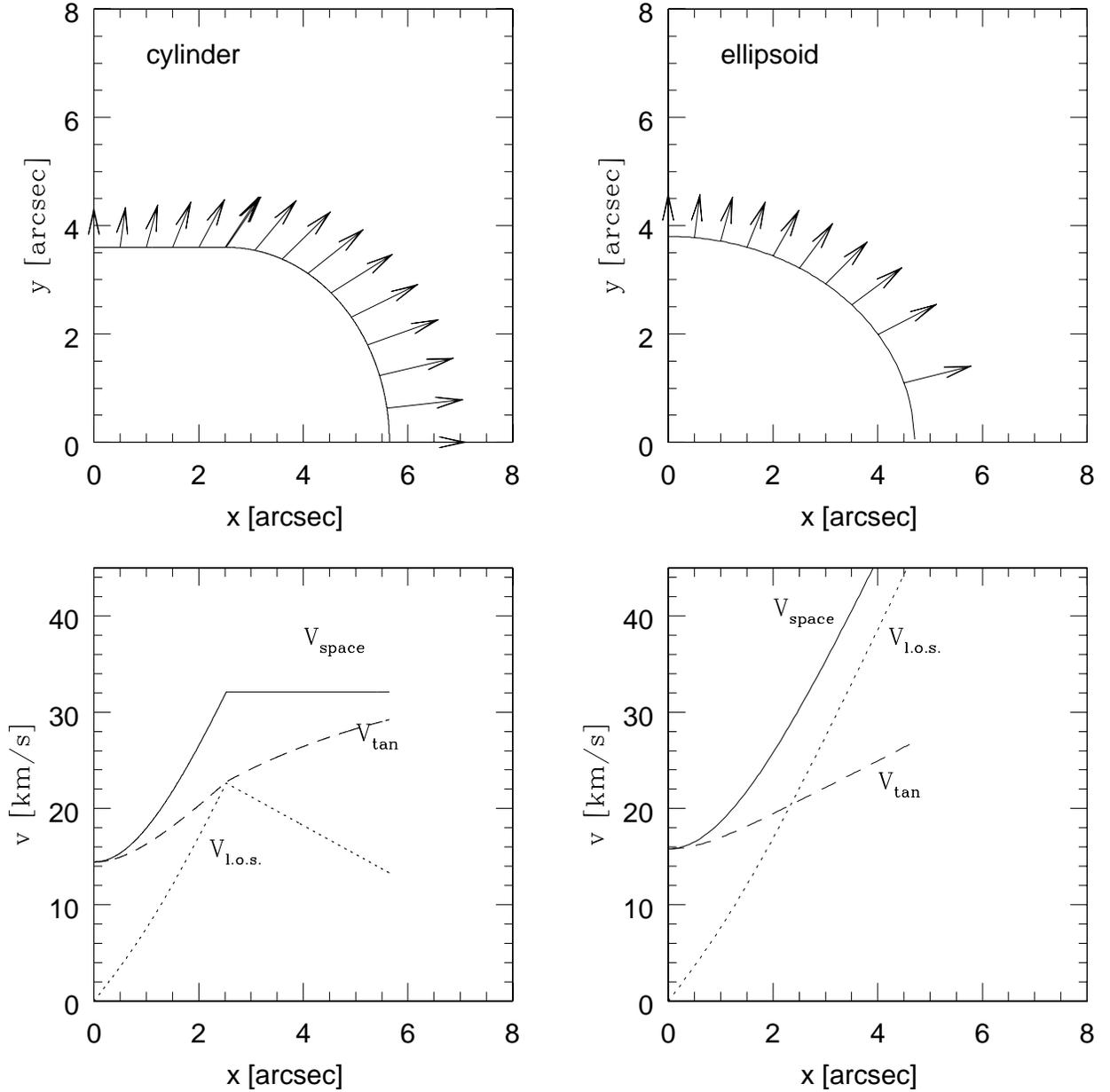}
 \caption{\label{cylinder.ps} The cylinder and ellipsoid model of NGC
7027. Both models have an inclination with the line of sight of 30
degrees. The upper panels show the direction and magnitude of the
tangential expansion.  The bottom panels illustrate the velocities
(space, tangential and line of sight) of the outer edge of the
observed nebula.} 
\end{figure*}

To represent this morphology, following \citet{Masson1989}, we first
assume a cylindrical model for the nebula, with $r_{\rm minor}=3.6$
arcsec and $r_{\rm major}\cos i=5$ arcsec, observed on an inclination
of $i=30$ degrees with respect to the line of sight.  The axial ratio
of the unprojected cylinder is 1.4. (The observed elongation is
slightly amplified by the thickness of the cylinder.) This model,
together with the velocity field above, predicts that the surface area
of the nebula increases by a fraction of $0.0020 /(D \rm [kpc])\,
yr^{-1}$.  We also considered the ellipsoid model
of \citet{Masson1989} (their model 3).  This requires a much larger
intrinsic axial ratio and therefore reaches much higher velocities in
the polar direction, assuming the velocity field above.  But the
tangential expansion is very similar and we find a fractional increase
of surface area of $0.0021 /(D \rm[kpc])\, yr^{-1}$, i.e. almost
identical to the cylindrical model.  The tangential expansion
predicted from the two models is illustrated in
Fig. \ref{cylinder.ps}.

The kinematical models compared to the measured increase of
$0.00251\pm0.00015\,\rm yr^{-1}$, yield a distance of $ D = 820\times
f_c\,\rm pc$. Here, $f_c$ covers several correction factors which
still need to be applied: the decline in surface brightness
\citep{Kawamura1996}, and most importantly, the difference in speed
between the velocity of the gas and the speed of the ionization front
\citep{Masson1989}. We find that the assumption of constant $T_b$ 
underestimates the expansion by 10
percent.\footnote{\citet{Terzian1997} notes that the correction will
also depend on the presence of clumps, filaments, etc. which may show
a different surface brightness decline from the diffuse gas.}  The
photo-ionization model shows that the ionization front moves 40 per
cent faster than the gas.  This is the most uncertain part of
expansion distances, and in extreme cases the ionization front can
move up to three times faster than the gas
\citep{Schoenberner2005}. The structure of the ionization front
requires a time-dependent model, which we have not attempted.

Combining the two factors, which act in opposite direction, yields
$f_c=1.25$. For comparison, \citet{Masson1989} finds a factor of
$(1/0.87)=1.15$. The difference can be taken as an indication for the
uncertainty in the models. This uncertainty dominates over our uncertainty on
the angular expansion. It will be difficult to improve on the
expansion distance without better nebular models.

Taking the average correction factors found by us and
\citet{Masson1989} (i.e. $f_c = 1.20$), we derive an expansion distance to 
NGC\,7027 of $d=980\pm100$\,pc.

\subsection{Previous expansion distance measurements}

\begin{table*}
\caption[]{\label{expdist}  NGC 7027  expansion distances derived from
 previous papers. The column ``Corrected distance'' incudes additional correction factors to the
 original results, to allow comparison with our distance. $\Theta$ is the 
angular diameter.}
\begin{tabular}{lllll}
\tableline
\tableline
  $\dot\Theta$     &  Vexp  &  Distance  &   ref. & Corrected distance \\
%    [marcsec/yr]   &   [km/s] &    [pc]      &        & [pc] \\
    $\left[ \rm mas/yr\right]$  &   [km/s] &    [pc]      &        & [pc] \\
\tableline
 2D &            3D &    $980\pm100$            &    This paper  & \nodata \\                    
$4.7 \pm 0.7$  &  21  &  $940\pm200$  &   \citet{Masson1986} &  \nodata \\
$4.2 \pm 0.6$ & $17.5\pm1.5$ &   $880\pm150$  &   \citet{Masson1989}
 & $ 970\pm 165$  \\
$5.25 \pm 0.55$ &  $17.5\pm1.5$   &  $703\pm95$   &   \citet{Hajian1993} 
     & $880\pm120$ \\
$5.25 \pm 0.55$ &   18.9    &   $760$      &   \citet{Walsh1997}
      & 950  \\
$4.2 \pm 0.06$ &  $13\pm1$       & $ 650 \pm100$    & \citet{Bains2003} 
      & $710\pm110$\\
\tableline
\end{tabular}
\end{table*}

Previous distance measurements are listed in Table \ref{expdist}.  Two
independent expansion measurements are reported in \citet{Hajian1993}
and \citet{Masson1989}, together with an intermediate result reported in
\citet{Masson1986}. The other papers re-used their measurements.

\citet{Masson1989} applied several corrections to the measured expansion,
to obtain the bulk gas motion from the measured expansion of the
ionization front. One of these corrections is not warranted, as they
include a decline of the 4.9\,GHz radio flux which we do not
confirm. Removing this correction decreases the change in angular
diameter $\dot \Theta$ by 0.4\,mas\,yr$^{-1}$.  They also derive a
correction for the difference in the velocity of the ionization front
and the gas, and include the correction for the declining brightness
temperature. \citet{Bains2003} simply corrected the result
of \citet{Masson1989} for a lower expansion velocity, and so
implicitly used the other correction factors. \citet{Hajian1993} used
the measured expansion as the bulk gas motion, but ignored the
differential movements; they also did not correct for inclination
effects.  \citet{Walsh1997} present a careful determination of the 3D
velocity field, based on optical velocity cubes, and apply this to the
inclined ellipsoidal model of \citet{Masson1989} to obtain the
velocity in the plane of the sky. They use the expansion measurement
of \citet{Hajian1993} but do not correct for the ionization front
velocity or the brightness temperature decrease.

Comparing the values in Table \ref{expdist}, the low expansion velocity of
\citet{Bains2003} is noticable. Their velocity is based on [O\,{\sc iii}]
which measures velocities in an intermediate excitation region closer to the
central star. Lines formed near the ionization front in planetary
nebulae, such as [N\,{\sc ii}], commonly show higher velocities
\citep[e.g.][]{Gesicki2003}.

Table \ref{expdist} reports corrected distances, where we include 'missing' 
correction factors, and for \citet{Masson1989} remove the correction for
the flux decrease.  The resulting distances  cluster around 950\,pc.

\section{Evolution and mass of the central star}

\begin{figure*}
 \includegraphics[width=14cm]{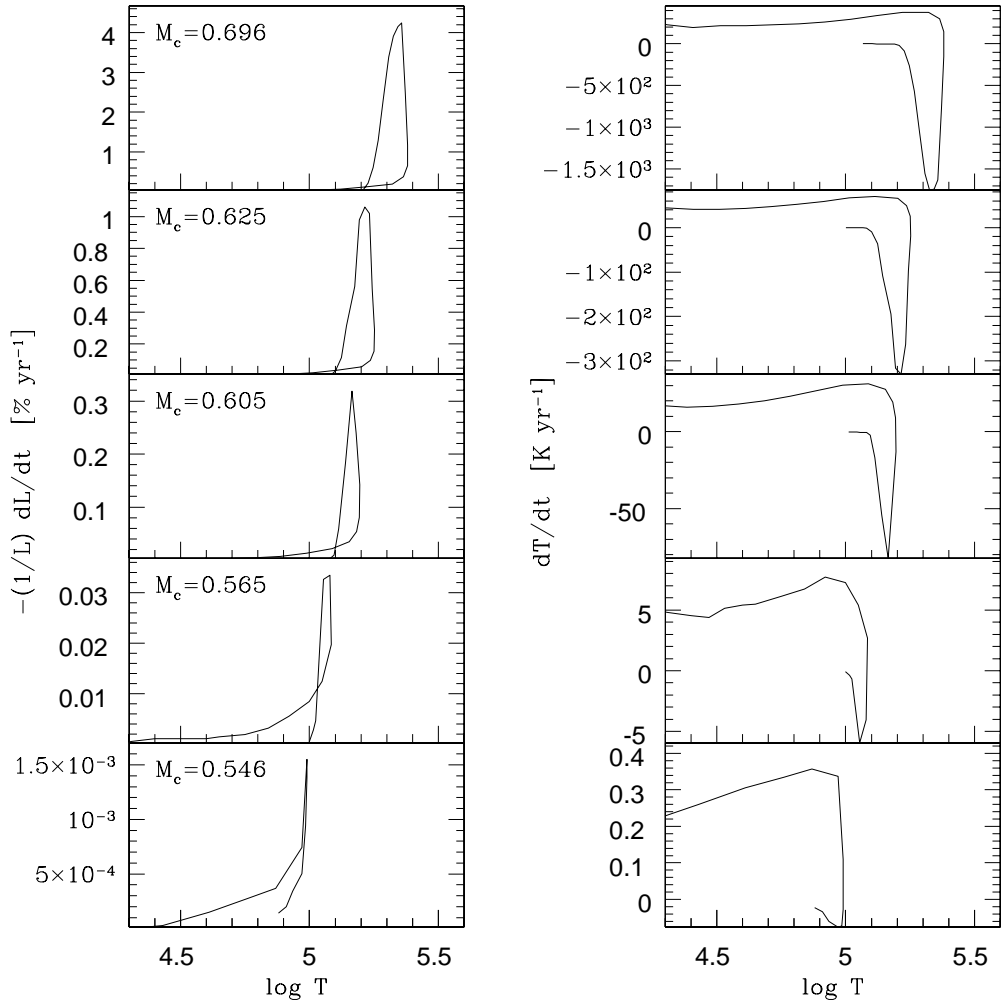} \caption{\label{mass}
%{tl_change.ps}
The rate of change in temperature and luminosity, for 
different stellar masses, on the Bl\"ocker tracks. }
\end{figure*}

The radio measurements reveal a clear decrease in the flux at higher
frequencies, where the emission is optically thin. The radio flux in
this regime is proportional to the recombination rate within the ionized
nebula. Recombination time scales are fast since the nebular densities
are high. The Cloudy model indicates a recombination time of 4\,yr. We
can therefore assume ionization equilibrium such that the
recombination rate is equal to the ionization rate.

\begin{figure}
 \includegraphics[width=9cm]{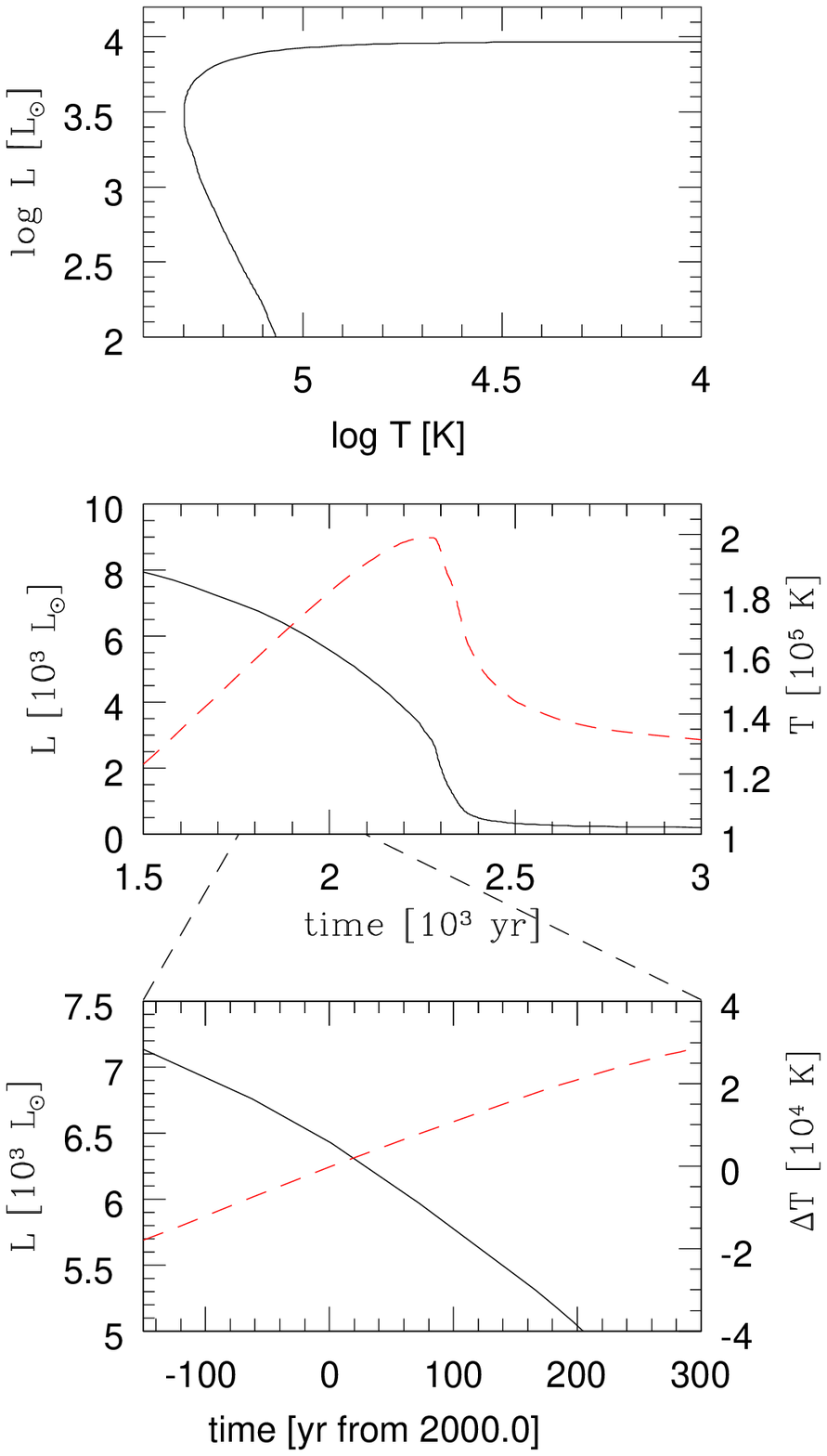} \caption{\label{time}An
illustration of the expected evolution of the central star. Here we
use an interpolated Bl\"ocker track of mass 0.664\,M$_\odot$. The top
panel shows the track in the HR diagram. The middle panel shows the
luminosity (solid line) and temperature (dashed line)
as a function of time, where we defined the time zeropoint such that 
time 2000 correspond to the current parameters of the star.
The bottom panel zooms in
on the time corresponding to the current temperature of the central
star of NGC\,7027.  In this panel, the temperature $T$ is plotted
as the difference between $T(t)$ and $T(0)$.  }
\end{figure}

The ionization rate is determined by the number of ionizing photons, $Q({\rm
  H})$, where every photon gives one ionization. The decline of the optically
thin flux indicates a reduction in the number of ionizing photons. The model
described above gives a fractional decline of $1/Q\,dQ(\rm H)/dt = -0.00145
\pm 0.00005\,\rm yr^{-1}$. The number of ionizing photons depends on the
stellar temperature and luminosity.

To get the evolution of the stellar temperature and luminosity with time, we
use the evolutionary tracks from \citet{Bloecker1995}. These are calculated
for hydrogen-burning central stars (a minority of stars may be helium burners,
in particular the [WC] stars). The tracks are available for a number of
different masses: 0.546, 0.565, 0.605, 0.625, 0.696, 0.836, and
0.940\,M$_\odot$. The initial masses are much higher; the stars lose between
20 and 80 per cent of their mass on the Asymptotic Giant Branch. The
initial-final mass relation remains uncertain. The rates of change of
temperature and luminosity for different final masses are plotted in
Fig.~\ref{mass}.

We have interpolated between the individual Bl\"ocker models to obtain a
much finer core-mass resolution, and to determine for each core mass
the relation between ($L(t)$, $T_{\rm eff}(t)$). The procedure is
described in \citet{Frankowski2003}, and we used tracks obtained from
K. Gesicki (priv. comm.).

Fig. \ref{time} illustrates the expected evolution of the star for one of
these interpolated tracks. On the post-AGB track, the temperature increases
while the remaining hydrogen envelope is reduced by burning. Once the nuclear
burning ceases, a fast drop in luminosity occurs, typically by a factor of 100
over a century. The temperature reaches its maximum during the initial decline
and afterwards begins a slow decline. This evolution gives rise to the
characteristic 'knee' in the HR diagram. The maximum temperature which is
reached depends on the mass of the star. The timescales before the knee, in
terms of the temperature increase per year, are a strong function of stellar
mass. This in principle allows for an accurate measurement of this mass.
Recent papers have derived the evolutionary time scale of a PN from the
expansion age of the nebula, assuming that the heating started immediately
after the nebula was originally ejected. This yields a very narrow mass
distribution for the central stars, between 0.58 and 0.65\,M$_\odot$ with a
peak at $0.61\,$M$_\odot$ \citep{Gesicki2007}. This is based on indirect
measurements of the stellar heating time. NGC\,7027 allows us to apply the
same technique using a direct measurement of the current rate of evolution.

$Q$(H) depends on stellar temperature and luminosity. For sufficiently hot
stars it decreases with increasing temperature (assuming constant luminosity)
because each photon carries more energy. In order to derive a relationship
between $T_{\rm eff}$ and $Q(H)$, we need to know the spectral energy
distribution of the star. We used 2 assumptions: a blackbody spectrum and
Rauch H-Ni line-blanketed NLTE model atmospheres with log($g$) = 7 or 8
\citep{Rauch2003}. The observed decrease in $Q(\rm H)$ would, for constant
luminosity and an initial temperature of $T_{\rm eff} = 1.61 \times 10^4$\,K
\citep{Beintema1996}, translate into a temperature change of $dT/dt = 273\pm
17\,\rm K\,yr^{-1}$ using the Rauch model atmosphere
calibration. Using the interpolated Bl\"ocker tracks, we find that
this rate of temperature evolution is reached for $M = 0.675 \,\rm
M_\odot$. In practice, this yields an upper limit to the mass. The
luminosity is also decreasing, and this contributes to the rate of
evolution. Thus, the real temperature increase is less than this
limiting value. For each mass, we computed the temperature at which
$1/Q\,dQ(\rm H)/dt = -0.145\%\,\rm yr^{-1}$. For each track there are
two solutions, one on the horizontal track and one on the cooling
track. The latter can be rejected, as the luminosity at this point on
the cooling track is $\log L/L_\odot \approx 2.4$ while the observed
value is close to $\log L/L_\odot = 4$ \citep{Middlemass1990, Beintema1996,
Wolff2000}.

\begin{figure}
  \includegraphics[width=9cm]{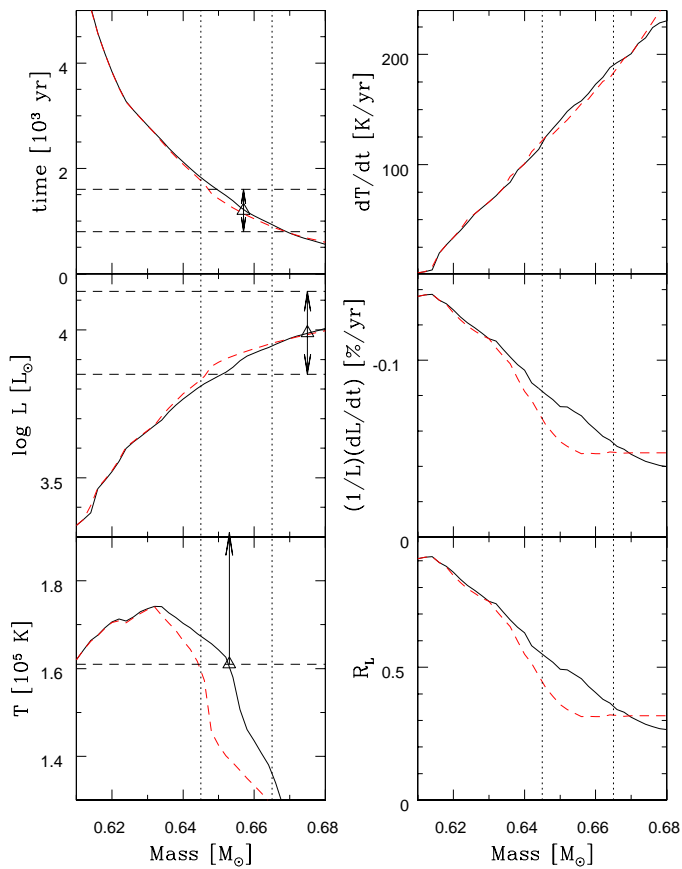} \caption{\label{dQ}For each
    stellar mass, the left panels show the temperature, luminosity and
    age for which the models reproduce the observed $1/Q\,dQ/dt$. The
    solid line was derived using a blackbody calibration for $Q(H)$,
    while the dashed line was derived using the Rauch model atmosphere
    calibration. The top right panel shows the change in temperature,
    and the middle-right panel the fractional change luminosity, as
    function of mass. The bottom right panel shows the fractional
    contribution $R_L$ of the luminosity change to the total
    $1/Q\,dQ/dt$. The luminosity decrease dominates the evolution at
    lower masses while the temperature increase is more important at
    higher masses.}
\end{figure}

The observed $1/Q\,dQ/dt$ can be reproduced for a range of masses, as
illustrated in Fig.~\ref{dQ}. Hence we need an additional constraint
to break the degeneracy. For this we use the kinematical age of the
nebula. We calculate the kinematical age using the
method of \citet{Gesicki2000}: it uses the observed expansion velocity
and radius, but measures these quantities at 0.8 of the outer radius.
PN velocity fields show strong acceleration near the outer boundary,
due to the overpressure of the ionized gas. The value of 0.8 of the
outer radius is closer to the mass-averaged radius, and gives more
stable values. \citet{Gesicki2000} also apply a correction to allow
for the acceleration of the nebula, which we did not use as NGC\,7027
is still ionization bounded and the acceleration is still minor. Using
the velocity field above (eq.~\ref{vfield}), we find a dynamical age
of 1200\,yr.  The uncertainty on the method is estimated as $\pm
400$\,yr, including the distance uncertainty.  The top panel of
Fig. \ref{dQ} shows the corresponding age of the star on the Bl\"ocker
track, measured from the start of the post-AGB phase. The horizontal
dashed lines indicate the age uncertainty.

The luminosity at the time the specific value for $1/Q\,dQ/dt$ is reached, for
each mass, is plotted in the middle-left panel of Fig. \ref{dQ}. The
luminosity corresponding to our kinematical age is log $L/L_\odot$ = 3.86 --
3.91, depending on the calibration for $Q(H)$. The luminosity is derived by
\citet{Wolff2000} based on HST photometry of the star: they give a range of
values, depending on the uncertain extinction. Converting their values to our
preferred distance gives $\log L/L_\odot$ = 3.89 -- 4.09. The
\citet{Beintema1996} model, scaled to our distance, gives 
$\log L/L_\odot =4.03$. The luminosity corresponding to our
kinematical age agrees with the lower end of the range found by
\citet{Wolff2000}. Our distance uncertainty of 100\,pc corresponds to
an uncertainty in $\log L/L_\odot$ of 0.084, and this is sufficient to
bring the predictions and observations in agreement.

The bottom left panel of Fig. \ref{dQ} shows the stellar temperature for which
the observed $1/Q\,dQ/dt$ is reached. For the adopted $T_{\rm eff} = 1.61
\times 10^4$\,K \citep{Beintema1996}, there are two solutions, one at
$\sim$0.61\,M$_\odot$ and one at $\sim$0.644 -- 0.652\,M$_\odot$ depending on
the $Q(H)$ calibration. The lower solution is clearly ruled out by our
constraint on the kinematical age, but the higher value is in fair agreement
(especially if the blackbody calibration is adopted). There is however
disagreement in the literature for the stellar temperature: published values
range from $T_{\rm eff} = 1.40\times10^5$\,K \citep{Middlemass1990} to $2.19
\times 10^5$\,K \citep{Zhang2005}, while our revised model also igves a somehwat higher temperature. The first detection of the [Ne\,{\sc vi}]
and [Ar\,{\sc vi}] lines by \citet{Beintema1996} favors a high
temperature, and makes the low temperature found by
\citet{Middlemass1990} seems doubtful. But none of the higher values
can be ruled out at this stage, so that we indicate the adopted value
as a lower limit.  We find no solution in the Bl\"ocker tracks for
temperatures above $\sim 1.75 \times 10^5$\,K: for the indicated mass
range, such high temperatures are only reached in the 'knee', when the
luminosity drops too fast. Higher mass stars reach such temperatures
while still on the horizontal track, but their temperatures increase
too fast. Hence we can summarize that the large uncertainty on the
stellar temperature makes this a very poor constraint on the core
mass. At this stage we cannot even confirm that the temperature is in
agreement with the theoretical tracks. More observations and/or
modeling are needed to settle this point.

We can summarize the previous discussion as follows. The kinematical
age of the nebula is consistent with a stellar mass of $M = 0.655\,\rm
M_\odot$, while the luminosity favors a more massive star but given
the distance uncertainties may still be consistent with this
mass. Masses below 0.645\,M$_\odot$ appear to be inconsistent with the
luminosity or age.  The stellar temperature favors a mass below
0.66\,M$_\odot$. Combining these requirements, we adopt a
mass derived from the Bl\"ocker tracks of

\begin{equation}
  M_{\rm Bl} = 0.655\pm 0.01 \,\rm M_\odot.
\end{equation}

\noindent For this mass and assuming a blackbody atmosphere, we 
find

\begin{eqnarray}
\nonumber  \frac{d T}{dt}  &=& 155\pm35 \,\rm K\, yr^{-1} \\
  \frac{1}{L} \frac{dL}{dt} &=& -0.070\pm0.015 \,\rm \% \, yr^{-1}
\end{eqnarray}

\noindent  The individual contributions from the changing luminosity
and temperature are indicated in the right panels of
Fig. \ref{dQ}. The bottom right panel  shows the fraction of the 
$1/Q\,dQ/dt$ which can be attributed to the change in luminosity, $R_L$. 
Approximately 45-65\%\ of the decline in the number of ionizing photons is 
due to the increasing temperature, and the remaining 55--35\%\ to a decreasing
luminosity. The individual evolutionary rates of the luminosity
and temperature are indicated in the right panels of
Fig. \ref{dQ}. For lower masses, the luminosity decrease dominates the
total $1/Q\,dQ/dt=-0.145$, while for higher masses, the temperature change is
more important.

\subsection{Model uncertainty}

The derivation above yields a consistent, well defined mass as derived
from the interpolated Bl\"ocker tracks. The error margin does not,
however, include any systematic uncertainty in the interpolated Bl\"ocker
models. This uncertainty becomes evident when comparing to the
alternative tracks of \citet{Vassiliadis1994}.

The tracks of \citet{Vassiliadis1994} (hereafter VW95) differ from the
Bl\"ocker tracks mainly in the treatment of the mass loss. They assume an
early and fast end to the AGB mass loss, leading to an extended period between
the end of the AGB and the onset of ionization, the so-called transition time.
After this transition, the VW94 models speed up and during the PN phase, the
model tracks behave qualitatively similar. The shorter transition times of the
Bl\"ocker models provide a better fit to the ages of planetary nebulae.

\begin{figure}
  \includegraphics[width=9cm]{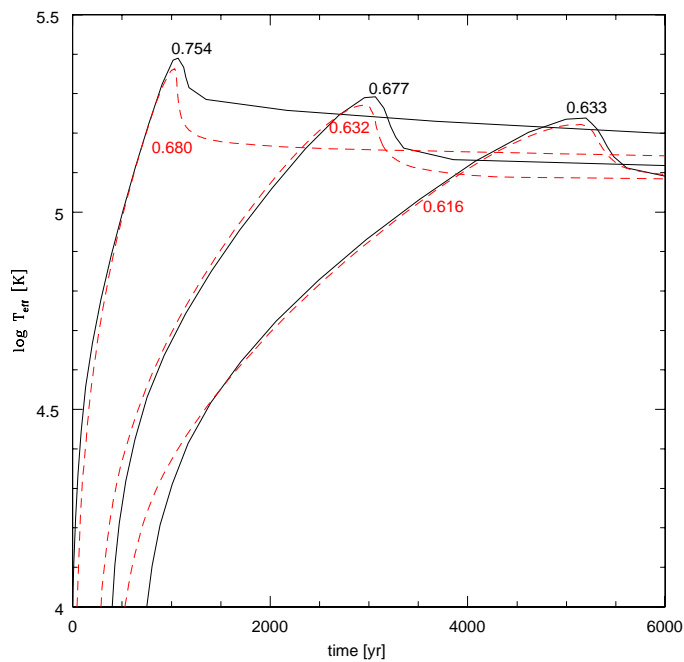} \caption{\label{VW}The solid
lines show three tracks from the \citet{Vassiliadis1994} models; the
dashed lines show interpolated Bl\"ocker models with the same
heating rate. The solid lines are shifted in the x direction to force a
fit.  The tracks are labeled by their stellar mass.}
\end{figure}

We have compared the heating rates of the VW94 models to those of Bl\"ocker.
This shows a surprising difference, in that the former show a slower
temperature increase for the same mass, even for hot, ionizing central stars
where the mass loss through the stellar wind is unimportant. For the same
heating rate, the VW94 models require a higher stellar mass. This is
illustrated in Fig. \ref{VW}, where for three tracks of VW94 we show the
interpolated Bl\"ocker track with the same heating rate, labeling both tracks
by their masses. To take account of the different transition times, the VW94
tracks are shifted horizontally to force overlap with the corresponding
Bl\"ocker track. For the Bl\"ocker mass range derived for NGC\,7027, the
tracks differ by $\sim 0.04$\,M$_\odot$. Thus, using the VW94 tracks we would
have derived a mass of around 0.7\,M$_\odot$.

We have used the Bl\"ocker tracks because of their better treatment of
the early post-AGB phase. Higher mass models show a higher luminosity,
and allow for a higher stellar temperature. This may indicate that
better agreement could have been obtained with the VW94 models.
However, it is not currently possible to decide between the models and
new model calculations would be warranted.

This large difference appears to be related to the relation
between the envelope mass and the stellar temperature. Comparing the
$M_{\rm env}$--$T_{\rm eff}$ relation in Fig. 4 of
\citet{Bloecker1995}, to those derived from the VW94 models as shown
in \citet{Frankowski2003}, shows that the VW94 models of the same core
mass and envelope mass are much hotter, or for the same temperature
have a higher envelope mass. The explanation of this discrepancy is
not clear to us,.

The dominant source of uncertainty in the derived mass is therefore
a systematic one, related to the adopted stellar evolution model.

\subsection{Initial--final mass relation}
The initial-final mass relation is not well
known, but is best constrained from white dwarfs in open clusters.
A comprehensive study is presented by \citet{Ferrario2005}, re-analyzed by
\citet{Williams2007}. The latter proposes the following linear relation, based 
on white dwarfs in open clusters:

\begin{equation}
 M_{\rm f} = (0.132\pm0.017)M_{\rm i} + (0.33\pm0.07)
\end{equation}

\noindent For NGC\,7027, the nebular abundances indicate a progenitor mass in
the range 3-4\,M$_\odot$ \citep{Salas2001}. Assuming the lowest mass,
the final mass should be $>0.726$\,M$_\odot$.  The (Bl\"ocker) mass
determinations place NGC\,7027 well below the proposed initial-final mass
relation. A mass as high as implied by the relation above can be
excluded, as such a massive star would have evolved much faster than
observed. The mass implied from the VW94 models is closer to the
proposed relation, although still below its lower limit. 

However, individual white dwarfs within the same cluster show a very
large range of masses. For the two clusters with initial masses around
3\,M$_\odot$, Praesepe and Hyades, 4 out of 11 stars have masses below
0.7\,M$_\odot$, while 3 have masses $>0.8$\,M$_\odot$. A mass of
0.65\,M$_\odot$ for NGC\,70727 would be unexceptional, and for
instance identical to that of 0431+125, with a progenitor mass of
2.8\,M$_\odot$ and a white dwarf mass of $M_f =0.652 \pm 0.032\,\rm
M_\odot$.

We conclude that the derived mass of the central star of NGC\,7027
does agree with the initial-final mass distribution, assuming a
progenitor mass near 3\,M$_\odot$. This distribution is however not
well represented by the proposed single, linear
relation. \citet{Ferrario2005} also discuss a higher order relation
which may be in better agreement with  at least the lower masses.

\section{Radio flux density calibrators}
\label{baars_revised}

The radio flux scale in use at the VLA is based on \citet{Baars1977},
who determined a scale based on absolute flux density measurements of
the young supernova remnant Cas A and the extragalactic radio source
Cygnus A.  These two sources are uniquely suitable for absolute
measurements, as they provide typically 1000 Jy of spectral radio flux
density.  The flux density of Cas A has been determined for the range
0.3--30\,GHz. Cas A is itself evolving, and is decreasing in flux by
$0.98\pm0.04 -0.30*\log \nu$ per cent per year, where $\nu$ is the
frequency in GHz.

The rapid evolution, which may not be entirely uniform, makes Cas A
less useful as a time-independent calibrator. More importantly, both
Cas A and Cygnus A are far too extended for use by modern radio
interferometers.  \citet{Baars1977} therefore transferred the scale to
several small-diamater (typically less than 1'') extragalactic
sources.  For the purposes of setting the flux scale, we utilize the
Baars expression for 3C\,295, which we believe to be non-varying as it
is an extragalactic radio emitter with physical size scales in excess
of 1\,kpc.  The only compact radio emission comes from the nucleus,
which for this source constributes less than 5 mJy -- or less than 1\%
of the total at any frequency.  The effect of any secular variations
on the total can thus be neglected.  As mentioned earlier, the
variation of the ratio of 3C\,286 to 3C\,295 is less than 0.01\% per year,
thus justifying our assumption that 3C\,286 is also constant.  The
original (Baars) flux scale adopted for 3C\,286 is listed in Table
\ref{flux_scale}.

The Cloudy model discussed above showed the need for time-independent
correction factors which dramatically improved the quality of the
fit. The factors are listed in Table \ref{modelflux}. Although
imperfections in the model may be present, it is unlikely that these
cause the observed deviations, and we believe they originate in the
adopted flux scale.  The model for NGC\,7027 cannot improve on the
{\it absolute} flux scale, but can be used to test for its consistency
over the full frequency range.

\citet{Baars1977} gives uncertainties  of 2 per cent for the measurements 
of Cas A at the frequency range discussed here, and $\sim5$\%, or
worse, for that of the secondary calibrators. The discrepancies we see
are of order 1 per cent, and therefore consistent with their accuracy.
Under the assumption that the correction factors represent real
inaccuracies in the flux density scale, the bottom row of Table
\ref{flux_scale} gives the revised values for the flux densities of
3C\,286.

NGC\,7027 is a valuable source for establishing the flux density scale
for two reasons.  It is a bright, relatively compact source which is
observable over a wide range of frequencies. Second, the frequency
dependence and the temporal evolution of its emission are well
understood.  These make it possible to calculate the flux of NGC\,7027
for any frequency within the range covered here (1.2--45\,GHz), and
for any time over which our linear evolutionary model remains
valid. This makes NGC\,7027 uniquely suitable for use in calibrating
non-standard radio frequencies, such as the 30\,GHz receivers which
have recently come into operation.

\begin{table}
\caption[]{\label{flux_cal} Flux densities for NGC\,7027, for various
frequency bands currently in use, $b$ is the secular variation of the flux density.
%The last column gives, for NGC\,7027, the change of the flux 
%in mJy/year.
  }
\begin{flushleft}
\begin{tabular}{llllllllll}
\tableline
\tableline
Band & Freq. [GHz] & Flux [mJy] & $b$ [mJy/yr] \\
            &       & (2000.0)  & \\  
\tableline
{\it VLA} \\
 LL  &    1.465  &  1542.8 &       $+$4.00 \\
 CC  &    4.885  &  5521.5 &       $-$3.88 \\
 XX  &    8.435  &  5929.4 &       $-$7.12 \\
 UU  &   14.965  &  5812.5 &       $-$7.99 \\
 KK  &   22.485  &  5604.8 &       $-$7.95 \\
 QQ  &   43.315  &  5203.8 &       $-$7.51 \\
{\it VLA} \\
 HH  &    1.430  &  1470.5 &       $+$3.89 \\
 18  &    1.665  &  1966.9 &       $+$4.49 \\
 VC  &    4.985  &  5552.3 &       $-$4.06 \\
 VX  &    8.465  &  5929.8 &       $-$7.13 \\
 VU  &   15.315  &  5802.4 &       $-$8.00 \\
 VK  &   22.235  &  5611.1 &       $-$7.96 \\
 VQ  &   43.135  &  5206.4 &       $-$7.51 \\
\tableline
{\it OCRA} &  30       & 5433.8 &       $-$7.79 \\
{\it CBI}  &  31       & 5413.6 &       $-$7.77 \\
{\it VSA}  &  33       & 5375.0 &       $-$7.72 \\
\tableline
\end{tabular}
\end{flushleft}
\end{table}

To illustrate this, Table \ref{flux_cal} lists the
parameters for the radio flux of NGC\,7027 for  common
frequency bands used at the VLA, as well for several new,
30\,GHz experiments. These were calculated using the derived
photo-ionization structure and the temporal evolution of 
optical depth and number of ionizing photons. Gaunt factors were
calculated with high precision for each of these bands.

Using the same technique, Table \ref{spec_full} gives the full
model radio spectrum of NGC\,7027. 

The linear extrapolation from epoch 2000.0 becomes invalid over very
long time scales. Thus at 1.275 GHz, the slope changes from +2.87
mJy/yr in 1981.0 to +3.52 mJy/yr in 2010.0. Extrapolating a linear
evolution from 1981.0 predicts a 2010.0 flux of 1094.7\,mJy, versus an
actual model flux of 1187.8\,mJy, an error of almost 1\%. At 5\,GHz the
error over 29 years is 3.5\,mJy (0.06\%), while  at 9\,GHz it is
down to 1.3\,mJy (0.02\%).

The linear approximation to the flux changes therefore works well,
but the errors become noticeable especially at the lowest frequencies.
%One can instead  recalculate the 
%flux for each frequency and epoch, and this may be required where the
%full accuracy of the model is required. An on-line calculator is
%available on \verb+http://www.pa.uky.edu/~peter/flux/+.

\begin{table}
\caption[]{\label{spec_full} Model flux densities for NGC\,7027, 
on a logarithmic frequency scale. The full (finer) version of this
table is available by request to the author. %electronically.
  }
\begin{flushleft}
\begin{tabular}{rrr}
\tableline
\tableline
 Freq. [GHz] & Flux [mJy] & $b$ [mJy/yr] \\
               & (2000.0)  & \\  
\tableline
  1.000000 &  692.40  & $+$2.043   \\
  1.995262 & 2662.17  & $+$4.554   \\
  3.019952 & 4303.79  & $+$1.423   \\
  3.981072 & 5120.87  & $-$1.796   \\
  5.011872 & 5560.25  & $-$4.112   \\
  6.025596 & 5773.05  & $-$5.519   \\
  7.079458 & 5879.05  & $-$6.427   \\
  8.128305 & 5923.46  & $-$6.996   \\
  9.120108 & 5934.69  & $-$7.347   \\
 10.000000 & 5930.05  & $-$7.561   \\
 15.135612 & 5807.58  & $-$7.998   \\
 20.417379 & 5658.59  & $-$7.989   \\
 30.199517 & 5429.71  & $-$7.787   \\
 40.738028 & 5242.74  & $-$7.561   \\
 50.118723 & 5110.67  & $-$7.387   \\
\tableline
\end{tabular}
\end{flushleft}
\end{table}

\section{Conclusions}
We have presented the results of a 25-year monitoring program of
NGC\,7027 at radio wavelengths. These results provide conclusive
evidence for on-going evolution of this object. At low frequencies,
where the nebula is optically thick, the radio flux is increasing
approximately linearly with time, at a rate of $0.251\pm0.015\%\,\rm
yr^{-1}$. This is shown to be directly caused by expansion of the
nebula. From the observed increase and the known velocity field, we
derive an expansion distance of $d=980\pm100\,$pc. The distance
includes several correction factors: the most important one of these
is the difference between the velocity of the gas and the velocity of
the ionization front. The uncertainties in these correction factors
dominate the final result: improved measurements of the rate of
expansion will therefore not a-priori yield better distances.

At higher frequencies, where the nebula is optically thin, the data
shows a steady decline of the radio flux, at a rate of
$-0.145\pm0.005\%\,\rm yr^{-1}$.  We show that this can only be caused
by evolution of the exciting star.  The number of ionizing photons is
dropping, due to an {\it increase} in stellar temperature and a
decrease in stellar luminosity.

Evolutionary models of post-AGB stars are used to fit the observed
decrease in ionizing photons. Using interpolated evolutionary tracks of
\citet{Bloecker1995}, we find a well-constrained mass
for the central star of NGC\,7027, of $0.655\pm0.01\,\rm M_\odot$.
However, the alternative models of \citet{Vassiliadis1994} evolve
slower, and would indicate masses higher by about 0.04\,M$_\odot$.
The discrepancy is related to a difference in the relation between
envelope mass and effective temperature. For the Bl\"ocker model, we
find a temperature increase of $135\pm35 \,\rm K\, yr^{-1}$, and a
luminosity decrease of $- 0.075\pm0.025 \,\rm \% \, yr^{-1}$.  The
mass is consistent with those of white dwarfs in open clusters with
initial masses around 3\,M$_\odot$, or slightly below.

The photoionization model of NGC\,7027 reveals  small residuals
which are frequency-dependent but constant with time. Correcting
for these gives a large improvement on the fit. We propose that they
are caused by inaccuracies in the relative fluxes of 3C286, 
based on the Baars flux scale. If this is correct, it is possible to
improve the internal consistency of the Baars scale. A set of revised
flux densities is proposed for 3C\,286 which takes this correction into
account. Using the models for NGC\,7027 developed in this paper,
we  calculate its time-dependent radio flux densities for the standard
continuum bands used at the VLA, as well as some high frequency
bands used by a number of new experiments.

\section*{Acknowledgements}

PvH acknowledges support by the Belgian Science Policy Office under
grant MO/33/017. The Very Large Array (VLA) is operated by the
U.S. National Radio Astronomy Observatory, a facility of the National
Science Foundation operated under cooperative agreement by Associated
Universities, Inc. AAZ received support from an STFC rolling grant,
and also gratefully acknowledges the hospitality of the SAAO where part
of this work was carried out.

{\it Facilities:} \facility{VLA}

\bibliographystyle{mn2e}

\bibliography{N7027}

\end{document}